\pdfoutput=1
\documentclass[reprint,aps,prd,superscriptaddress,amsmath,amssymb,nobibnotes,longbibliography,nofootinbib,floatfix]{revtex4-2}

\usepackage{graphicx}

\usepackage{amsfonts}
\usepackage{needspace}

\usepackage[dvipsnames]{xcolor}
\usepackage[normalem]{ulem}

\usepackage[colorlinks=true,
            linkcolor=blue,
            urlcolor=blue,
            citecolor=blue,
            anchorcolor=blue]{hyperref}
\hypersetup{
  pdftitle={Action-selected currents and a singular conservation closure in two-fluid energy--momentum-squared cosmology},
  pdfauthor={Ozgur Akarsu, Bilal Bulduk, Nihan Katirci},
  pdfsubject={Action-selected currents, a singular conservation closure, and exact backgrounds in two-fluid energy--momentum-squared cosmology},
  pdfkeywords={energy--momentum squared gravity, conservation closure, interacting fluids, rank degeneracy, exact cosmology}
}

\usepackage{orcidlink}

\begin{document}

\title{Action-selected currents and a singular conservation closure in
two-fluid energy--momentum-squared cosmology}

\author{\"{O}zg\"{u}r Akarsu\,\orcidlink{0000-0001-6917-6176}}
\email{akarsuo@itu.edu.tr}
\affiliation{Department of Physics, Istanbul Technical University, 34469 Maslak, Istanbul, T\"{u}rkiye}

\author{Bilal Bulduk\,\orcidlink{0000-0002-7446-9640}}
\email{bulduk21@itu.edu.tr}
\affiliation{Department of Physics, Istanbul Technical University, 34469 Maslak, Istanbul, T\"{u}rkiye}
\affiliation{Department of Physics, Gebze Technical University, 41400 Gebze, Kocaeli, T\"{u}rkiye}

\author{Nihan Kat{\i}rc{\i}\,\orcidlink{0000-0002-9492-3791}}
\email{nkatirci@dogus.edu.tr}
\affiliation{Department of Electrical and Electronics Engineering,
Do\u{g}u\c{s} University, 34775 \"{U}mraniye, Istanbul, T\"{u}rkiye}

\begin{abstract}
In multi-fluid matter-type gravity, constituent currents follow from the
off-shell matter action or an additional closure, not from the Bianchi
identity alone. We study a spatially flat two-fluid background inspired
by scale-independent energy--momentum squared gravity (EMSG), with
species-separable
$f=\alpha\sum_i\sqrt{T_{\mu\nu(i)}T_{(i)}^{\mu\nu}}$ excluding mixed
contractions. We assume a vanishing contracted-Hessian contribution
and separately conserved total conventional and modification
sectors; neither is derived here from a fluid action, nor does
the minimal separable action select the latter, so the model is an
effective closure, not microscopic EMSG. Specified constant-$w$
Brown and regular Schutz actions give separate conservation at nonzero
gravitational weights. For unequal constant equations of state and every
$\alpha\ne0$, this closure instead yields
Barrow--Clifton-type exchange with $\alpha$-independent coefficients
set by each modification partner's pressure mismatch with its parent's
equation of state. Hence, except for the zero-exchange stiff--vacuum
pair, the $\alpha\to0$ limit generically fails to recover uncoupled
general relativity, a singular behavior of the closure, not of EMSG. The
modified Li\'enard equation for $H$ admits exact parametric integration.
We prove that the discriminant governing rank loss of the
density-reconstruction map is positive for all finite $w_1\ne w_2$;
at the rank-degenerate couplings the gravitating source has one
constant equation of state on admissible backgrounds, hence constant
deceleration. For distinct $w_i\in(-1,1)$, the maximal density-positive
interval has logarithmic width fixed by $(w_1,w_2)$; imposing $H^2>0$
can shorten the cosmological interval. Generic backgrounds have evolving
deceleration; a dust--vacuum example crosses into acceleration, but its
positive-density branch cannot reach an early matter era.
No observationally viable model is claimed.
\end{abstract}

\maketitle
\section{Introduction}
\label{sec:intro}

Matter-type gravity modifies the material part of the Einstein--Hilbert action
through functions such as $f(\mathcal L_{\rm m})$~\cite{Harko:2010mv},
$f(\mathcal T)$~\cite{Harko:2011kv}, and
$f(T_{\mu\nu}T^{\mu\nu})$~\cite{Katirci:2013okf,Roshan:2016mbt,
Akarsu:2017ohj,Board:2017ign}, the last of which is commonly called
energy--momentum squared gravity (EMSG); see Ref.~\cite{Cipriano:2024jng} for a
recent overview. Here
$\mathcal T=g^{\mu\nu}T_{\mu\nu}$ denotes the trace of the energy--momentum
tensor (EMT). At the level of the metric equations these models can be recast
as general relativity (GR) sourced by an enlarged effective matter sector
\cite{Akarsu:2023lre}. Once a complete off-shell matter action and a particular
split between ``conventional'' and ``modification'' terms have been fixed,
metric variation fixes the two partial tensors and their relative current.
That split is not invariant under a reshuffling of terms in the total matter
Lagrangian, and it does not by itself select every possible current assignment
when several material species are present.

Interacting-fluid cosmologies provide the complementary motivation. At the
homogeneous level one usually postulates a scalar transfer kernel $Q$ in the
continuity equations. Linear choices proportional to $H\rho_i$ are analytically
tractable and widely studied \cite{Barrow:2006hia,Caldera-Cabral:2008yyo,
Chimento:2009hj,Kaeonikhom:2022ahf}; broader reviews are given in
Refs.~\cite{Bolotin:2013jpa,Wang:2016lxa,Wang:2024vmw}. Such a background
kernel is not, however, a complete covariant model: its four-vector completion,
momentum transfer, perturbations, and stability depend on additional physics
\cite{Valiviita:2008iv,Faraoni:2014vra,Tamanini:2015iia}. This distinction is
particularly important here because the closure considered below is imposed at
the level of conservation laws rather than derived from new interspecies terms
in a matter action. 

Action-based formulations of interacting relativistic fluids illustrate what
is additionally required for a microscopic completion
\cite{CarrilloGonzalez:2017cll,Iosifidis:2024ksa,Koivisto:2015qua}.
In the scalar--fluid example of Ref.~\cite{Koivisto:2015qua}, the
interaction is specified in the Lagrangian and the background equations,
linear perturbations, and large-scale-structure predictions follow from
that action. This is an example of a completion, not a realization of the
particular prescription and closure used here.
A conceptually related prescription in which ordinary
and effective modification stresses are conserved separately was considered in
an $f(R,\mathcal T)$ setting in Ref.~\cite{Moraes:2016mlp}; the construction
below differs by resolving two conventional species and their modification partners
into four sub-sectors.

Late-time reconstructions of an interacting vacuum first indicated a data
preference for an exchange that switches on at low
redshift~\cite{Salvatelli:2014zta}, and phenomenological dark-sector
interactions have since been tested extensively against the $H_0$ and
structure-growth tensions
\cite{DiValentino:2017iww,Yang:2018euj,DiValentino:2019ffd,Hoerning:2023hks},
while sign-changing kernels have been confronted with both earlier and
DESI-era data~\cite{Cai:2009ht,Wei:2010cs,Giare:2024smz,Li:2025owk,Sabogal:2025mkp,Silva:2025hxw} and the kernel has been
reconstructed model-independently, with mild hints of an oscillatory,
direction-changing transfer~\cite{Escamilla:2023shf}; see
Ref.~\cite{CosmoVerseNetwork:2025alb} for a broader review.
Analyses including DESI DR2 have renewed interest in interacting
models. For example, Ref.~\cite{Pan:2025qwy} finds improved minimum
$\chi^2$ for most dataset combinations, while information criteria and
Bayesian evidence remain inclined toward $\Lambda$CDM.%
 These developments motivate interacting-fluid studies without
validating the specific background closure or supplying the constituent
currents considered here.%

The question pursued here is therefore narrower than constructing a new
interacting-dark-energy fit. Given an algebraic EMSG stress tensor, which
features of a two-fluid background are selected by the action and which are
consequences of an additional conservation assignment? This distinction
matters because a diagonalized coupling can appear ``theory fixed'' even when
its value is fixed only conditionally on the closure, and because an apparently
weak-coupling limit can be singular if the coupling has been divided out of a
conservation equation. Exact Friedmann--Lema\^itre--Robertson--Walker (FLRW) solutions provide a direct diagnostic of
both effects.

For two species, the total Bianchi identity supplies one collective zero-sum
condition for four sub-sectors: the two conventional EMTs and their two induced
modification partners. A minimal separable and independently
diffeomorphism-invariant matter action naturally conserves each species together
with its own partner. Other partitions are not selected by the Bianchi identity
alone and require either additional matter interactions or a phenomenological
closure. We use \emph{current} for a covariant four-vector divergence and
reserve \emph{kernel} for the scalar source term obtained after projection and
algebraic diagonalization of the background equations.

Our case study adopts
\begin{equation}
f=\sum_{i=1}^{2}\alpha_i
\sqrt{T_{\mu\nu(i)}T_{(i)}^{\mu\nu}},
\end{equation}
sets $\alpha_1=\alpha_2$, and imposes separate conservation of the total
conventional and total modification sectors. This sector-separation rule is
compatible with the total Bianchi identity, but it is not the on-shell
conservation law selected by the minimal separable action. The construction is
therefore an EMSG-structured phenomenological closure. We also state explicitly
the algebraic perfect-fluid metric-variation prescription used to evaluate the
EMSG tensor; the on-shell identities $\mathcal L_{\rm m}=p$ and
$T_{\mu\nu}=(\rho+p)u_\mu u_\nu+pg_{\mu\nu}$ do not determine that off-shell
prescription uniquely.
This prescription yields the algebraic EMSG stress commonly used in
cosmological applications~\cite{Akarsu:2023lre}, which is why its consequences
are isolated here.

These are independent restrictions: species separability specifies which
matter invariants are retained, excluding mixed contractions such as
$T_{\mu\nu(1)}T_{(2)}^{\mu\nu}$ that a generic function of the total
$T_{\mu\nu}T^{\mu\nu}$ would contain (Sect.~\ref{sec:iii}); the
metric-variation prescription specifies the algebraic modification stress;
and sector separation specifies its conservation assignment. Bianchi
compatibility alone establishes none of the three. In particular, the
subsequent interaction coefficients are conditional on their combination,
rather than consequences of the EMSG action alone, and are not claimed for a
generic two-fluid EMSG theory. Table~\ref{tab:attribution} summarizes this
attribution.%

Conditional on these assumptions, the diagonalized conventional densities obey
a Barrow--Clifton-type system. Its two coefficients are fixed by the species
equation-of-state (EoS) parameters rather than fitted as independent couplings, although this
fixing is a property of the full closure and not of the minimal EMSG action
alone. The densities share two eigenmodes, and the resulting Hubble rate obeys
a modified Li\'enard equation. Independently, the matrix
that reconstructs the densities from the gravitational equations becomes rank
deficient at special values of $\alpha$. Those values are modal-cancellation
subfamilies, not additional dynamical branches. We derive the generic and
rank-deficient solutions and state the finite intervals on which all
conventional densities remain positive.

The linear interacting-fluid modes and the associated power-law and de Sitter
expansion laws have substantial precedent
\cite{Barrow:2006hia,Caldera-Cabral:2008yyo,Chimento:2009hj}, and related
scale-independent EMSG backgrounds were studied in Ref.~\cite{Akarsu:2023nyl}.
The species-separable continuity system, its species-dependent coefficient map,
the modified Li\'enard equation, and several exact backgrounds appeared
previously in B.B.'s public M.Sc. thesis~\cite{Bulduk:2024thesis}. The present
work reassesses their theoretical status and extends the analysis in four
respects. First, it separates action-selected conservation laws from
Bianchi-compatible phenomenological closures, using explicit constant-$w$
Brown and Schutz actions as a comparison with the algebraic prescription.%
 Second, it identifies the
sector-separated $\alpha\ne0$ family as a closure whose modification-sector
constraint disappears at $\alpha=0$, shows that its $\alpha$-independent
exchange kernel is set by the pressure mismatch between each modification
partner and its parent fluid, and that its generic nonzero-exchange
solutions fail to approach uncoupled GR; the stiff--vacuum zero-exchange
exception is kept explicit.%

Third, it interprets the special values of $\alpha$ as rank loss in
the density-reconstruction map, shows that at these couplings the
gravitating source obeys a single constant equation of state, and proves
positivity of the associated discriminant for every finite
unequal-equation-of-state pair.
Fourth, besides the vacuum- and stiff-fluid endpoints, it proves that for
distinct $w_i\in(-1,1)$ every maximal density-positive interval has a
fixed logarithmic width; the additional condition $H^2>0$ can shorten
the admissible FLRW interval.%
 These results delimit precisely which statements follow
from the algebraic background prescription and which would require a complete
off-shell matter action.

The novelty is thus the consistency analysis and conceptual reclassification,
together with the global discriminant proof and physical-domain analysis, not
the linear interacting-fluid ansatz, the existence of two density modes, or
the already known elementary expansion laws.%

The scope is deliberately restricted to exact homogeneous backgrounds. Because
the sector-separation closure has no microscopic matter-action realization in
the present work, perturbation equations derived for other EMSG prescriptions
cannot simply be imported. A completion must either derive the constituent
currents and their perturbations from an explicit off-shell action or specify
the covariant transfer four-vectors, including their momentum components,
perturbations, and phenomenological closure relations, directly. 
Background kinematics can already be tested, but a complete stability
analysis and observational inference involving cosmological perturbations
require one of these routes to be fixed. In particular, the constant-$q$
solutions are local analytic diagnostics, not proposed complete alternatives
to $\Lambda$CDM; Sect.~\ref{sec:kinematics} distinguishes their limitations
from the time-dependent deceleration of the general solution.%

Section~\ref{sec:EI} sets out the action-selected tensor structure, and
Sect.~\ref{sec:iii} distinguishes covariant conservation assignments from
background kernels. Section~\ref{sec:casestudy} introduces the scale-independent EMSG case
study. Generic cosmological solutions are presented in Sect.~\ref{sec:cossol},
and the rank-deficient subfamilies are analyzed in
Sect.~\ref{sec:alphaparticular}. Section~\ref{sec:conc} summarizes the results and
their limitations.

\section{Action-selected conservation laws in matter-type gravity}
\label{sec:EI}
Matter-type gravity models can be generated by adding a function
$f(\boldsymbol{\theta}_A)$ of a matter scalar $\boldsymbol{\theta}_A$ to the
matter sector of the Einstein--Hilbert action. We work on a four-dimensional
Lorentzian spacetime and restrict attention to scalars constructed locally from
the matter fields and the metric, without explicit matter--curvature
contractions. Examples are the matter Lagrangian itself,
$\theta_A=\mathcal{L}_{\rm m}$, the EMT trace
$\theta_A=\mathcal{T}=g^{\mu\nu}T_{\mu\nu}$, and the quadratic
contraction $\theta_A=T_{\mu\nu}T^{\mu\nu}$. The corresponding
action is
\begin{equation}
\mathcal{S}=\int {\rm d}^4x \sqrt{-g}\,\left[\frac{1}{2\kappa}R+\mathcal{L}_{\rm m}+f(\boldsymbol{\theta}_A)\right],
\label{actgen}
\end{equation}
where $\kappa=8\pi G$, with $G$ Newton's gravitational constant. We use units
$\hbar=c=1$.

Assume that the metric enters $\mathcal L_{\rm m}$ algebraically, with no
metric-derivative dependence; dependence on the matter fields and their
derivatives is implicit. The conventional EMT is then
\begin{align}
   \label{gen-tmunudef}
T_{\mu\nu}&=-\frac{2}{\sqrt{-g}}\frac{\delta(\sqrt{-g}\mathcal{L}_{\rm m})}{\delta g^{\mu\nu}}=\mathcal{L}_{\rm m}g_{\mu\nu}-2\frac{\partial\mathcal{L}_{\rm m}}{\partial g^{\mu\nu}}.
\end{align}
The metric equations can be represented as Einstein equations sourced by an
enlarged matter sector by defining \cite{Akarsu:2023lre}
\begin{equation}
\mathcal{L}_{\rm m}^{\rm tot}=\mathcal{L}_{\rm m}+f(\boldsymbol{\theta}_A).
\end{equation}
This rewriting does not make the theory physically equivalent to GR with the
original uncoupled matter model: the total matter Lagrangian and its matter
equations have changed. Moreover, the partial tensors defined below depend on
the chosen split of $\mathcal L_{\rm m}^{\rm tot}$ into $\mathcal L_{\rm m}$
and $f$; a reshuffling changes their individual currents while leaving the
total EMT invariant.
The total EMT defined by Eq.~\eqref{gen-tmunudef} is
\begin{align} \label{emt-tot}
T_{\mu\nu}^{{\rm tot} }=-\frac{2}{\sqrt{-g}}\frac{\delta(\sqrt{-g}\mathcal{L}_{\rm m}^{{\rm tot} })}{\delta g^{\mu\nu}}=T_{\mu\nu}+T_{\mu\nu}^{{\rm mod} }.
\end{align}
Variation of Eq.~\eqref{actgen} then gives
\begin{align}  
\label{modfieldeq}
G_{\mu\nu}=\kappa T_{\mu\nu}+\kappa T_{\mu\nu}^{{\rm{mod}} },
\end{align}
where $G_{\mu\nu}=R_{\mu\nu}-\frac{1}{2}Rg_{\mu\nu}$ is the Einstein tensor and $T_{\mu\nu}^{\rm{mod}}$ denotes the new terms coming from the metric variation of $\sqrt{-g} \,f(\boldsymbol{\theta}_A)$ and reads
\begin{align}
T_{\mu\nu}^{\rm mod}&=-\frac{2}{\sqrt{-g}}\frac{\delta\left[\sqrt{-g}\,f(\boldsymbol{\theta}_A)\right]}{\delta g^{\mu\nu}}\nonumber\\
&=f(\boldsymbol{\theta}_A)g_{\mu\nu}-2\sum_A f_A(\boldsymbol{\theta}_A)\,\Theta_{\mu\nu}^{(A)},
\end{align}
where
\begin{equation}
    f_A(\boldsymbol{\theta}_A)
\equiv
\frac{\partial f(\boldsymbol{\theta}_A)}{\partial\boldsymbol{\theta}_A},\qquad\Theta_{\mu\nu}^{(A)}\equiv\left.\frac{\partial\boldsymbol{\theta}_A}{\partial g^{\mu\nu}}\right|_{\psi},
\end{equation}
and $\psi$ collectively denotes the matter variables held fixed in the metric
variation. We assume
$\boldsymbol{\theta}_A=\boldsymbol{\theta}_A(g^{\mu\nu},\psi)$ with no
derivatives of the metric.

One can gather related new tensors as
\begin{align}
\label{NT-Lm}
\Theta_{\mu\nu}^{(\mathcal{L}_{\rm m})}&\equiv\frac{\partial\mathcal{L}_{\rm m}}{\partial g^{\mu\nu}}=\frac{1}{2}\left(\mathcal{L}_{\rm m}g_{\mu\nu}-T_{\mu\nu}\right),\\
\label{NT-T}
\Theta_{\mu\nu}^{(\mathcal{T})}&\equiv
\frac{\delta\mathcal{T}}{\delta g^{\mu\nu}}
=\mathcal{L}_{\rm m}g_{\mu\nu}-T_{\mu\nu}
-2g^{\alpha\beta}\frac{\partial^2\mathcal{L}_{\rm m}}
{\partial g^{\mu\nu}\partial g^{\alpha\beta}},\\
\Theta_{\mu\nu}^{(T^2)}&\equiv\frac{\delta\left(T_{\alpha\beta}T^{\alpha\beta}\right)}{\delta g^{\mu\nu}}\label{NT-T2}\\
&=2T_{\mu}^{\lambda}T_{\nu\lambda}-\left(\mathcal{T}+2\mathcal{L}_{\rm m}\right)T_{\mu\nu}+\mathcal{L}_{\rm m}\mathcal{T}g_{\mu\nu}\nonumber\\
&\quad-4T^{\alpha\beta}\frac{\partial^2\mathcal{L}_{\rm m}}{\partial g^{\mu\nu}\partial g^{\alpha\beta}}.\nonumber
\end{align}
 Taking the covariant derivative of Eq.~\eqref{modfieldeq} with the twice contracted Bianchi identity, $\nabla^{\mu} G_{\mu\nu}=0$, we obtain
\begin{align}
\nabla^\mu T_{\mu\nu}^{\rm tot}=\nabla^\mu\left(T_{\mu\nu}+T_{\mu\nu}^{\rm mod}\right)=0,
\end{align}
which does not imply conservation of either partial tensor separately. The
resulting exchange is a decomposition of the conserved total tensor; the
modification term need not represent an independent propagating field. We write
\begin{equation}
\label{eq:int}
\nabla^{\mu}T_{\mu\nu}\equiv-\mathcal{Q}_{\nu} \quad \textnormal{and} \quad \nabla^{\mu}T_{\mu\nu}^{{\rm mod} }\equiv+\mathcal{Q}_{\nu},
\end{equation}
where $\mathcal{Q}_{\nu}$ is the induced energy-transfer current. Its general
form is
\begin{align}
\mathcal{Q}_{\nu}&\equiv\nabla^\mu T_{\mu\nu}^{\rm mod}
\nonumber\\
&=\nabla_\nu f(\boldsymbol{\theta}_A)
-2\sum_A\Theta_{\mu\nu}^{(A)}\nabla^\mu f_A(\boldsymbol{\theta}_A)\nonumber\\
&\quad-2\sum_A f_A(\boldsymbol{\theta}_A)
\nabla^\mu\Theta_{\mu\nu}^{(A)}.\label{eq:intgen}
\end{align}
Thus, for a single material source with a complete off-shell action and a fixed
conventional/modification split, the corresponding current is fixed by metric
variation.\footnote{Here, the chain rule gives
\begin{align}
&\nabla_\nu f(\boldsymbol{\theta}_A)=\sum_A f_A(\boldsymbol{\theta}_A)\nabla_\nu\boldsymbol{\theta}_A\qquad\text{and}\nonumber\\
&\nabla^\mu f_A(\boldsymbol{\theta}_A)=\sum_B f_{AB}(\boldsymbol{\theta}_A,\boldsymbol{\theta}_B)\nabla^\mu\boldsymbol{\theta}_B,\nonumber
\end{align}
with
\begin{equation*}
f_{AB}(\boldsymbol{\theta}_A,\boldsymbol{\theta}_B)\equiv\frac{\partial^2f(\boldsymbol{\theta}_A,\boldsymbol{\theta}_B)}{\partial\boldsymbol{\theta}_A\partial\boldsymbol{\theta}_B}.
\end{equation*}}
The next section considers how a second species changes the admissible current
assignments.

\section{Bianchi-compatible conservation closures}
\label{sec:iii}
To obtain the four-sub-sector decomposition used below, we restrict the
two-species theory to a species-separable modification,
\begin{equation}
f(\boldsymbol{\theta}_{A(1)},\boldsymbol{\theta}_{A(2)})
=f_1(\boldsymbol{\theta}_{A(1)})+f_2(\boldsymbol{\theta}_{A(2)}).
\label{eq:separable-f}
\end{equation}
This is an additional model assumption, not a consequence of introducing two
matter species. In particular, a generic EMSG function constructed from the total tensor $T_{\mu\nu}=T_{\mu\nu(1)}+T_{\mu\nu(2)}$ contains mixed
contractions such as $T_{\mu\nu(1)}T_{(2)}^{\mu\nu}$, which are excluded by
Eq.~\eqref{eq:separable-f}.

For example, two comoving perfect fluids give
\begin{equation}
T_{\mu\nu(1)}T_{(2)}^{\mu\nu}
=\rho_1\rho_2+3p_1p_2
=(1+3w_1w_2)\rho_1\rho_2.
\label{eq:mixed-contraction}
\end{equation}
Writing $\overline T_{\mu\nu}\equiv T_{\mu\nu(1)}+T_{\mu\nu(2)}$ for the total
conventional tensor, one has
\begin{equation}
\begin{split}
\overline T_{\mu\nu}\overline T^{\mu\nu}
={}&T_{\mu\nu(1)}T_{(1)}^{\mu\nu}
+T_{\mu\nu(2)}T_{(2)}^{\mu\nu}\\
&+2(1+3w_1w_2)\rho_1\rho_2.
\end{split}
\label{eq:total-conventional-contraction}
\end{equation}
Consequently, for $w_1\ne w_2$ a square root of this total contraction is
not the sum of the two species square roots, even for a common coupling. A generic function of
the total contraction has joint species dependence and need not admit the
additive species-partner decomposition used below. The coefficient map and
closure-derived kernel in this paper therefore apply to the restricted
species-separable extension, not to a generic two-fluid EMSG theory.%

 We refer to each of the four tensors
\begin{equation}
T_{\mu\nu(1)},\qquad T_{\mu\nu(2)},\qquad T_{\mu\nu(1)}^{\rm mod},\qquad T_{\mu\nu(2)}^{\rm mod},
\end{equation}
as a sub-sector. A sub-sector labels the tensorial decomposition, whereas a
source below denotes a conventional density in the homogeneous background.
The total Bianchi identity constrains the sum of the four divergences but does
not authorize an arbitrary partition of them: a partition must follow either
from the matter-field equations or from an explicitly stated phenomenological
closure.

Separability ensures
$T_{\mu\nu}^{\rm mod}=T_{\mu\nu(1)}^{\rm mod}+T_{\mu\nu(2)}^{\rm mod}$,
so the total EMT admits the four-sub-sector representation
\begin{equation}
T_{\mu\nu}^{\rm tot}=T_{\mu\nu(1)}+T_{\mu\nu(2)}+T_{\mu\nu(1)}^{\rm mod}+T_{\mu\nu(2)}^{\rm mod}.
\end{equation}
Hence, the twice-contracted Bianchi identity is
\begin{equation}
\begin{aligned}
\nabla^\mu T_{\mu\nu}^{\rm tot}={}&
\nabla^\mu T_{\mu\nu(1)}+\nabla^\mu T_{\mu\nu(2)}\\
&+\nabla^\mu T_{\mu\nu(1)}^{\rm mod}
+\nabla^\mu T_{\mu\nu(2)}^{\rm mod}=0.
\end{aligned}
\label{eq:bianchi}
\end{equation}
We define an induced current for the divergence of each sub-sector,
\begin{align}
\nabla^\mu T_{\mu\nu(1)} &\equiv \mathcal{Q}_{\nu(1)},& \nabla^\mu T_{\mu\nu(2)} &\equiv \mathcal{Q}_{\nu(2)},
\label{eq:Q-matter-defs}\\
\nabla^\mu T_{\mu\nu(1)}^{\rm mod} &\equiv \mathcal{Q}_{\nu(1)}^{\rm mod},& \nabla^\mu T_{\mu\nu(2)}^{\rm mod} &\equiv \mathcal{Q}_{\nu(2)}^{\rm mod},
\label{eq:Q-mod-defs}
\end{align}
whose sum vanishes by the twice-contracted Bianchi identity,
\begin{equation}
\mathcal{Q}_{\nu(1)}+\mathcal{Q}_{\nu(2)}+\mathcal{Q}_{\nu(1)}^{\rm mod}+\mathcal{Q}_{\nu(2)}^{\rm mod}=0.
\label{eq:Q-total-zero}
\end{equation}
This is the collective condition imposed by the twice-contracted Bianchi
identity. For a single species, it reduces to Eq.~\eqref{eq:int},
with $\mathcal Q^{\rm mod}_{\nu(1)}=\mathcal Q_\nu$ and
$\mathcal Q_{\nu(1)}=-\mathcal Q_\nu$. The unindexed $\mathcal Q_\nu$
in Eq.~\eqref{eq:int} denotes the modification-sector divergence,
whereas the indexed currents in Eqs.~\eqref{eq:Q-matter-defs} and
\eqref{eq:Q-mod-defs} denote the divergences of their respective
sub-sectors. By itself, this collective condition does not
determine how the four sub-sector currents are paired or distributed.
Additional conservation assignments may follow from the matter-field
equations of a specified action or may be imposed phenomenologically.
The temporal component supplies total background conservation, including both
the conventional and modification sectors. To obtain separate equations for
$\rho_1(t)$ and $\rho_2(t)$, one must choose an action-selected assignment or
impose an additional closure and then solve the resulting relations for
$\dot\rho_1$ and $\dot\rho_2$:
\begin{align}
\dot{\rho}_1+3H(1+w_1)\rho_1&=\mathcal{Q}^{\rm D}_{(1)}(\rho_1,\rho_2;f_1,f_2),
\label{eq:diag-rho1}\\
\dot{\rho}_2+3H(1+w_2)\rho_2&=\mathcal{Q}^{\rm D}_{(2)}(\rho_1,\rho_2;f_1,f_2).
\label{eq:diag-rho2}
\end{align}
We call the four-vector divergences action-selected currents when they follow
from the matter equations of a complete action. Their temporal projections are background energy
transfer rates, while $\mathcal Q^{\rm D}_{(i)}$ denote the diagonalized
phenomenological kernels. The four sub-sector currents always obey the total
zero-sum condition. The sum
$\mathcal Q^{\rm D}_{(1)}+\mathcal Q^{\rm D}_{(2)}$ instead measures transfer
into the conventional pair and vanishes only for closures, such as sector
separation, that conserve that pair.

Following the standard covariant decomposition
\cite{Malik:2004tf,Clemson:2011an}, we write
\begin{equation}
    \mathcal{Q}_{\nu(i)}=\mathcal{Q}_{(i)}^{\rm U}u_\nu+F_{\nu(i)},
    \qquad u^\nu F_{\nu(i)}=0,
\end{equation}
where $\mathcal{Q}^{\rm U}_{(i)}$ and $F_{\nu(i)}$ are the energy- and
momentum-transfer parts relative to the total four-velocity $u_\nu$. When all
sub-sectors are retained, total conservation requires the respective sums of
both parts to vanish. Exact FLRW symmetry enforces
$F_{\nu(i)}=0$ at background order, but it does not determine the momentum
transfer at perturbative order. In comoving coordinates we use
$\mathcal{Q}_{0(i)}=-\mathcal{Q}^{\rm U}_{(i)}$; algebraic diagonalization then
produces the phenomenological scalar kernel $\mathcal{Q}^{\rm D}_{(i)}$.

\subsection{Bianchi-compatible conservation assignments}
The tensorial decomposition represents the covariant divergence of each
sub-sector by an individual four-vector, as defined in
Eqs.~\eqref{eq:Q-matter-defs} and \eqref{eq:Q-mod-defs}. The twice-contracted
Bianchi identity imposes only the collective zero-sum condition in
Eq.~\eqref{eq:Q-total-zero}; by itself, it does not determine how the four
currents are distributed. At this stage only the sub-sector currents are
defined. A two-density interaction kernel arises only after background
projection and algebraic diagonalization.

The examples below are representative, not exhaustive, conservation
assignments compatible with the collective Bianchi zero-sum condition. The
Bianchi identity alone selects none of them. A particular assignment may follow
from the on-shell matter-field equations of a specified microscopic action or
may instead be imposed as a phenomenological closure. The minimal separable
action realizes the species-paired assignment under the assumptions stated
below, whereas the sector-separated assignment used in the case study is adopted
as an additional closure. All subsequent cosmological results are conditional
on this choice.

Representative ways to organize the four currents are as follows.
\newcounter{qcounter}
\begin{list}
{\bfseries{}Species-paired~}
{
}
\item In this assignment, each conventional sub-sector exchanges energy--momentum only with its own modification partner:
\begin{align}
\mathcal{Q}_{\nu(1)}+\mathcal{Q}_{\nu(1)}^{\rm mod} &=0,
\label{eq:speciesi}\\
\mathcal{Q}_{\nu(2)}+\mathcal{Q}_{\nu(2)}^{\rm mod}&=0.
\label{eq:speciesii}
\end{align}
Thus, the two conventional sub-sectors do not exchange energy--momentum directly with each other. Instead, each conventional sub-sector and its own modification partner form a conserved pair.

For a minimal separable matter action in which the two matter components are independently diffeomorphism invariant and contain no direct interspecies coupling, the corresponding on-shell Noether identities imply
\begin{equation}
\nabla^\mu\left(T_{\mu\nu(i)}+T_{\mu\nu(i)}^{\rm mod}\right)=0,\qquad i=1,2.
\label{eq:minimal-onshell}
\end{equation}
Therefore, under these assumptions, the species-paired assignment is the
conservation structure naturally selected by the matter-field equations of the
minimal separable action.

\end{list}
\begin{list}
{\bfseries{}Sector-separated~}
{
\usecounter{qcounter}
}
\item Separate conservation of the conventional and modification sectors gives
\begin{align}
\mathcal{Q}_{\nu(1)}+\mathcal{Q}_{\nu(2)} &=0, 
\label{eq:sectori}\\
\mathcal{Q}_{\nu(1)}^{\rm mod}+\mathcal{Q}_{\nu(2)}^{\rm mod} &=0.
\label{eq:sectorii}
\end{align}
This is the sector-separation condition used in the main construction of the present work. It forbids net energy--momentum transfer between the conventional sector and the modification sector, but it does not require the individual sub-sectors to be separately conserved. This assignment is adopted here as a phenomenological closure; its realization through a complete microscopic matter-sector action is beyond the scope of the present work.

Unlike the species-paired assignment, the sector-separation conditions above do
not follow from the on-shell matter-field identities of the minimal separable
action described in Eq.~\eqref{eq:minimal-onshell}. In the present work, they
are imposed as an additional phenomenological closure. A fully covariant
microscopic realization would generally require further interspecies structure
or additional matter-sector interactions beyond the minimal separable setup.

\end{list}
\begin{list}
{\bfseries{}Cross-paired~}
{
\usecounter{qcounter}
}
\item In this case each conventional sub-sector is paired with the modification partner of the other sub-sector:
\begin{align}
\mathcal{Q}_{\nu(1)}+\mathcal{Q}_{\nu(2)}^{\rm mod} &=0,
\label{eq:crossi}\\
\mathcal{Q}_{\nu(2)}+\mathcal{Q}_{\nu(1)}^{\rm mod} &=0.
\label{eq:crossii}
\end{align}
This assignment represents a cross-coupled structure between the conventional and modification sectors. 

\end{list}
\begin{list}
{\bfseries{}One conventional sub-sector conserved~}
{
\usecounter{qcounter}
}
\item One may also impose that one conventional sub-sector is separately conserved. For example, taking the first conventional source to be separately conserved gives
\begin{align}
\mathcal{Q}_{\nu(1)} &=0,
\label{eq:grbasedi}\\
\mathcal{Q}_{\nu(2)}+\mathcal{Q}_{\nu(1)}^{\rm mod}+\mathcal{Q}_{\nu(2)}^{\rm mod}&=0.
\label{eq:grbasedii}
\end{align}
The analogous case with $1\leftrightarrow 2$ is obtained by separately conserving $\mathcal{Q}_{\nu(2)}$ instead. This assignment keeps one conventional source in its conventional conservation form, while its modification partner participates in the evolution of the remaining components.

\end{list}
\begin{list}
{\bfseries{}One modification sub-sector conserved~}
{
\usecounter{qcounter}
}
\item Similarly, one may impose that one of the modification sub-sectors is separately conserved:
\begin{align}
\mathcal{Q}_{\nu(1)}^{\rm mod} &=0,
\label{eq:modbasedi}\\
\mathcal{Q}_{\nu(1)}+\mathcal{Q}_{\nu(2)}+\mathcal{Q}_{\nu(2)}^{\rm mod}&=0.
\label{eq:modbasedii}
\end{align}
The analogous case with $1\leftrightarrow 2$ is obtained by separately conserving $\mathcal{Q}_{\nu(2)}^{\rm mod}$. 
\end{list}

\subsection{Representative background kernels}
The covariant currents describe energy--momentum exchange among the four
sub-sectors. Their homogeneous projections can be reorganized as effective
kernels for the two conventional densities. This reorganization is algebraic
and does not supply the missing matter action or perturbative momentum transfer.

For homogeneous perfect fluids with constant $w_i=p_i/\rho_i$, projection and
diagonalization of a specified conservation assignment give the local equations
\begin{align}
\dot{\rho}_1+3H(1+w_1)\rho_1&=\mathcal{Q}^{\rm D}_{(1)}(\rho_1,\rho_2;f_1,f_2),
\label{eq:intsysi}\\
\dot{\rho}_2+3H(1+w_2)\rho_2&=\mathcal{Q}^{\rm D}_{(2)}(\rho_1,\rho_2;f_1,f_2).
\label{eq:intsysii}
\end{align}
The quantities $\mathcal{Q}^{\rm D}_{(1)}$ and
$\mathcal{Q}^{\rm D}_{(2)}$ are phenomenological interaction kernels. They may
be obtained by projecting and diagonalizing a specified covariant assignment,
or postulated directly at background level. The representative covariant
assignments in Eqs.~\eqref{eq:speciesi}--\eqref{eq:modbasedii} and the
representative background forms below are therefore logically distinct.

The phenomenological kernels can be grouped as follows:

\begin{list}
{\bfseries{}Separate~}
{
}
\item The first possibility is
\begin{align}
\mathcal{Q}^{\rm D}_{(1)}=0\quad\text{and}\quad \mathcal{Q}^{\rm D}_{(2)}&=0.
\label{eq:standard}
\end{align}
Both conventional sources obey their conventional continuity equations.
\end{list}
\begin{list}
{\bfseries{}Modified effective EoS~}
{
}
\item The second possibility is that each diagonalized kernel depends only on its own density:
\begin{align}
\mathcal{Q}_{(1)}^{\rm D}=\mathcal{Q}^{\rm D}_{(1)}(\rho_1)\quad\text{and}\quad \mathcal{Q}^{\rm D}_{(2)}=\mathcal{Q}^{\rm D}_{(2)}(\rho_2).
\label{eq:modeos}
\end{align}
Provided \(H\rho_i\neq 0\), such a term may be absorbed into an effective EoS parameter,
\begin{equation}
w_i^{\rm eff}=w_i-\frac{\mathcal{Q}^{\rm D}_{(i)}}{3H\rho_i}.
\end{equation}
This changes the dilution law of each source without defining an exchange
between the two conventional densities; energy may still be redistributed among
the full set of sub-sectors.

\end{list}
\begin{list}
{\bfseries{}Symmetric exchange~}
{
}
\item The third possibility is a symmetric exchange between the two conventional densities:
\begin{align}
\mathcal{Q}^{\rm D}_{(1)}=\mathcal{Q}\quad\text{and}\quad \mathcal{Q}^{\rm D}_{(2)}=-\mathcal{Q}.
\label{eq:symmetric}
\end{align}
This is the usual form of an interacting two-fluid system at the
background equations level. A commonly used Barrow--Clifton-type linear kernel is
of the schematic form
\begin{equation}
\mathcal{Q}=3H\left(c_1\rho_1+c_2\rho_2\right),
\label{eq:linearkernel}
\end{equation}
with dimensionless $c_1$ and $c_2$
\cite{Barrow:2006hia,Caldera-Cabral:2008yyo,Chimento:2009hj}. We reserve
$\alpha$ and $\beta$ for the
EMSG and closure coefficients introduced below. In the present case the same
diagonal form arises only after the separable modification and the conservation
closure have both been specified.
\end{list}
\begin{list}
{\bfseries{}Asymmetric kernels~}
{
}
\item The fourth possibility is
\begin{align}
\mathcal{Q}^{\rm D}_{(1)}\neq 0,\quad \mathcal{Q}^{\rm D}_{(2)}\neq 0\quad\text{and}\quad \mathcal{Q}^{\rm D}_{(1)}&\neq \pm \mathcal{Q}^{\rm D}_{(2)}.
\label{eq:asymmetric}
\end{align}
This does not by itself contradict the total Bianchi identity. It means that
the two conventional densities, after diagonalization, do not form a closed
two-fluid exchange system by themselves. Part of the information carried by the
modification sector has been absorbed into the effective kernels
$\mathcal{Q}^{\rm D}_{(1)}$ and $\mathcal{Q}^{\rm D}_{(2)}$. Therefore
asymmetric diagonalized kernels should not be confused with a violation of
total energy--momentum conservation. Cross, conventional-based, and
modification-based current assignments can generically yield asymmetric or
mixed background kernels, but a current-balance assignment alone does not
determine a unique phenomenological class.

\end{list}

The labels above classify different properties and are not mutually exclusive.
The modified-EoS label identifies self-density terms that can be absorbed into
effective dilution parameters, whereas the asymmetric label characterizes the
relation between the two diagonalized kernels. A distinct cross-coupled case
requires explicit dependence of at least one kernel on the other conventional
density. In every case the complete set of four currents must satisfy
Eq.~\eqref{eq:Q-total-zero}; the diagonalized conventional kernels need not sum
to zero unless the conventional pair is itself conserved.

\section{Algebraic scale-independent EMSG with a singular sector-separation closure}
\label{sec:casestudy}
\nopagebreak[4]
The main background case study is conditional on both the algebraic
metric-variation prescription and the sector-separation closure specified
below; it is not the cosmology selected by the minimal separable EMSG
action. We keep a separate explicit action-based comparison to isolate
the consequences of these choices.%

We specialize to a spatially flat FLRW
background,
${\rm d}s^2=-{\rm d}t^2+a^2(t){\rm d}\boldsymbol{x}^2$. We use the on-shell
perfect-fluid representative $\mathcal L_{\rm m}=p$. In comoving coordinates
$\rho$ and $p$ are measured in the fluid rest frame, with
$u_\mu u^\mu=-1$, and the EMT is
\begin{align}
\label{em}
T_{\mu\nu}=(\rho+p)u_{\mu}u_{\nu}+p g_{\mu\nu},
\end{align} 
where $\rho>0$. The on-shell relation $\mathcal{L}_{\rm m}=p$ and the
perfect-fluid EMT~\eqref{em} do not determine the metric Hessian of the
off-shell matter Lagrangian. This ambiguity is not merely notational:
perfect-fluid representatives that are equivalent under minimal coupling can
define inequivalent theories when they enter a nonminimal matter--gravity
coupling~\cite{Faraoni:2009rk}. Within EMSG,
Ref.~\cite{Akarsu:2023lre} showed that setting the contracted-Hessian
contribution to zero produces a commonly used algebraic
energy--momentum-squared-field (EMSF) tensor but compromises the Lagrangian
formulation, whereas a thermodynamic evaluation retains a nonzero contribution
and changes the cosmological system~\cite{Marciu:2026cae}. Alternative recent
treatments of the off-shell fluid variation reach different conclusions about
the relevant metric variations and the role of the matter-Lagrangian %
representative~\cite{Haghani:2023uad,Boehmer:2026syd,VanKy:2026mdt,
Jayawiguna:2026das}. In a full Brown formulation the independent variables
include the particle-number flux, entropy, and constraint fields, while
Schutz's formulation uses velocity potentials and a thermodynamic pressure
function~\cite{Brown:1992kc,Schutz:1970my}. The metric dependence of the
derived density, pressure, and four-velocity must be retained when taking
successive variations; neither a constant background $w$ nor an on-shell
identity $\mathcal L_{\rm m}=p$ implies a vanishing contracted Hessian.

Recent trace-coupling calculations explicitly derive the metric and fluid
equations from the same off-shell action and check the consistency of their
conservation laws~\cite{Boehmer:2025afy}. Full Brown/Schutz analyses of
nonminimal fluid-Lagrangian couplings also find formulation-dependent
conservation and thermodynamic structures for the couplings they
study~\cite{Boehmer:2026syd}. These results concern their specified couplings,
not a derivation of the present EMSG tensor; their methodological relevance is
that an on-shell substitution cannot replace the complete variation.
We do not resolve the competing prescriptions in general. For the present
scale-independent modification, however, the explicit constant-equation-of-state
actions below provide a controlled comparison. The background case study
itself imposes%

\begin{equation}
T^{\alpha\beta}
\frac{\partial^2\mathcal{L}_{\rm m}}
{\partial g^{\mu\nu}\partial g^{\alpha\beta}}=0,
\label{eq:fluid-hessian-prescription}
\end{equation}
as a diagnostic algebraic prescription defining this background case study.
No Taub, Schutz, Brown, or other covariant fluid action
\cite{Taub:1954zz,Schutz:1970my,Brown:1992kc} realizing both
Eq.~\eqref{eq:fluid-hessian-prescription} and the generic interacting
current assignment is supplied. The explicit comparison below does not
provide such a completion, and no microscopic derivation of the case
study is claimed.%
 Nor does this prescription imply
that every component of the uncontracted Hessian vanishes. An action-level
completion would have to reproduce both this contracted quantity and the
adopted matter-current assignment through its metric and matter equations.
The EMSF tensor and exchange rates below are therefore prescription-dependent
background results, not microscopic action predictions.%

\paragraph{Comparison with explicit off-shell fluid actions.}
\label{sec:action-comparison}
The restriction is particularly transparent for two specified isentropic,
barotropic actions~\cite{Brown:1992kc,Schutz:1970my,Boehmer:2025afy}.
Here $p=w\rho$ is a constitutive identity with constant $w$, not merely a
relation imposed along a background solution. For Brown's densitized
number flux $J^\mu$, the full matter action can be written as
\begin{equation}
\begin{split}
S_{\rm B}&=\int\mathrm d^4x\left[-\sqrt{-g}\,\rho(n)+J^\mu V_\mu\right],\\
n&=\frac{\sqrt{-g_{\mu\nu}J^\mu J^\nu}}{\sqrt{-g}},
\qquad \rho(n)=\rho_\star n^{1+w},
\end{split}
\label{eq:brown-explicit-action}
\end{equation}
where $V_\mu=\partial_\mu\varphi+s\partial_\mu\vartheta+
\lambda_A\partial_\mu\xi^A$ contains the usual constraint fields, and
$\rho_\star>0$. The entropy $s$ is fixed in the isentropic sector.
In metric variation, $J^\mu$ and the constraint fields are held fixed;
the density $J^\mu V_\mu$ is metric independent and contributes no stress.
For the regular Schutz pressure action one instead holds the velocity
potentials fixed and uses
\begin{equation}
\begin{split}
S_{\rm S}&=\int\mathrm d^4x\sqrt{-g}\,p(\mu),
\qquad \mu=\sqrt{-g^{\mu\nu}V_\mu V_\nu},\\
\rho&=\mu p_{,\mu}-p,\qquad
p=w\rho,\qquad \rho=\rho_\star\mu^{(1+w)/w}.
\end{split}
\label{eq:schutz-explicit-action}
\end{equation}
The latter representation assumes $w\ne0,-1$ and an invertible
thermodynamic relation. With positive chemical potential and particle
number density, its ordinary-fluid branch has $w>-1$.
Brown's action includes dust, whereas dust is not obtained by substituting
$w=0$ into Eq.~\eqref{eq:schutz-explicit-action}. Vacuum is treated
separately by the constant Lagrangian $\mathcal L_{\rm v}=-\rho_{\rm v}$,
not by an invertible fluid chemical-potential relation.

In both specified actions the conventional stress has perfect-fluid form
off shell, so $T_{\mu\nu}T^{\mu\nu}=\sigma^2\rho^2$, where
$\sigma\equiv\sqrt{1+3w^2}$. For $\rho>0$, adding $f=\alpha\sigma\rho$
therefore gives
\begin{equation}
T^{\rm mod}_{\mu\nu}=\alpha k\,T_{\mu\nu},\qquad
k_{\rm B}=-\sigma,\qquad k_{\rm S}=\frac{\sigma}{w}.
\label{eq:action-stress-rescaling}
\end{equation}
This follows by varying the complete action densities: the Brown
modification is $-\alpha\sigma$ times the metric-dependent term, while
the Schutz modification is $\alpha\sigma/w$ times its pressure action.
The metric-independent density $J^\mu V_\mu$ does not affect this result:
if it is included in the scalar representative as $J^\mu V_\mu/\sqrt{-g}$,
its contributions to the explicit $\mathcal L_{\rm m}$ terms and to the
Hessian term of Eq.~\eqref{NT-T2} cancel identically, although it changes
the contracted Hessian itself.
A direct check gives
\begin{equation}
\begin{aligned}
\Theta^{(T^2),{\rm B}}_{\mu\nu}
 &= (1+w)\sigma^2\rho^2(g_{\mu\nu}+u_\mu u_\nu),\\
\Theta^{(T^2),{\rm S}}_{\mu\nu}
 &= -\frac{(1+w)\sigma^2\rho^2}{w}u_\mu u_\nu.
\end{aligned}
\label{eq:action-theta-comparison}
\end{equation}
For the specified Schutz scalar $\mathcal L_{\rm m}=p(\mu)$, the
contracted Hessian itself is
\begin{equation}
T^{\alpha\beta}
\frac{\partial^2\mathcal L_{\rm m}}
{\partial g^{\mu\nu}\partial g^{\alpha\beta}}
=\frac{1-w^2}{4w}\rho^2u_\mu u_\nu.
\label{eq:schutz-contracted-hessian}
\end{equation}
It is nonzero for generic regular $w$; it vanishes at $w=1$, as it also
does for the separately defined constant vacuum action at $w=-1$.
These direct variations provide a concrete comparison with the Hessian
sensitivity discussed in Refs.~\cite{Akarsu:2023lre,Marciu:2026cae}.

For species-separable versions of these actions, the Noether identity
\eqref{eq:minimal-onshell} reads
$(1+\alpha k_i)\nabla^\mu T_{\mu\nu(i)}=0$. On a regular branch with
$1+\alpha k_i\ne0$, it yields separate conservation and the GR dilution
laws. Thus the modification changes only the gravitational weights
$1+\alpha k_i$, i.e., the common factors multiplying $\rho_i$ and $p_i$ in
the field equations, not the constant-$w_i$ dilution exponents. If only sector separation is imposed
instead, then, for $\alpha\ne0$,
\begin{equation}
\mathcal Q_{\nu(1)}+\mathcal Q_{\nu(2)}=0,\qquad
k_1\mathcal Q_{\nu(1)}+k_2\mathcal Q_{\nu(2)}=0.
\label{eq:action-sector-current-system}
\end{equation}
For $k_1\ne k_2$ both currents vanish. For $k_1=k_2$ these conditions
coincide and the closure leaves exchange undetermined, whereas the
action-selected matter equations still enforce separate conservation
whenever $1+\alpha k_i\ne0$. In the Brown case, this exception $k_1=k_2$
occurs for distinct opposite equations of state, $w_1=-w_2$; since $k_{\rm S}$
is one-to-one on the regular Schutz domain, no such exception arises there.
Vanishing $1+\alpha k_i$ requires the original
action equations and is not covered by division by that weight.

The two formulations need not define the same nonminimally modified theory;
their gravitational weights differ, consistent with the formulation
dependence found for other nonminimal fluid couplings~\cite{Boehmer:2026syd}. Neither regular action-selected comparison reproduces
the generic nonzero exchange of the algebraic case study. This is not a
nonexistence theorem for other covariant actions with additional structure.
For constant-$w$ fluids, background effects of the scale-independent
modification obtained with the algebraic stress, such as the modified dilution
in Eq.~\eqref{eq:paired-algebraic-source}, therefore depend on that
prescription; in the explicit actions the modification changes only
gravitational weights.
Table~\ref{tab:attribution} summarizes these results together with
the algebraic-stress comparisons derived below.
We now return to the algebraic prescription
\eqref{eq:fluid-hessian-prescription} for the remainder of the case study.

 Under this prescription, substitution of
Eq.~\eqref{em} into Eq.~\eqref{NT-T2} for
$w=p/\rho={\rm const.}$ gives
\begin{equation} 
\label{eq:thetafrw}
\Theta_{\mu\nu}^{\rm EMSF}=-\rho^2(1+w)(1+3w) u_{\mu} u_{\nu}, 
\end{equation}
and the self-contraction of the EMT is
\begin{align}
T_{\mu\nu}T^{\mu\nu}=&\,\,\rho^2(3w^2+1).
\end{align}
Using Eq.~\eqref{eq:thetafrw} in Eq.~\eqref{eq:intgen}, the covariant current
for a perfect fluid is
\begin{equation}
\begin{aligned}
  \mathcal{Q}_{\nu}=&\nabla_{\nu}f+2(3w^2+4w+1)\left[\rho^2 u_{\mu} u_{\nu} \nabla^{\mu}f_{\mathbf{T}^2}\right. \\
  &\left.+f_{\mathbf{T}^2}\left(2\rho \nabla^{\mu}\rho \, u_{\mu} u_{\nu}+\rho^2\nabla^{\mu}\!\left(u_{\mu}u_{\nu}\right)\right)\right].
  \end{aligned}
\end{equation}
Only the temporal component contributes to an FLRW continuity equation.
Here $\mathcal Q_\nu$ is the single-source modification current of
Eq.~\eqref{eq:int}, not a conventional-species current in
Eq.~\eqref{eq:Q-matter-defs}. With $\mathcal Q_0=-\mathcal Q^{\rm U}$,
its projected scalar rate is
\begin{equation}
\begin{aligned}
\label{eq:creationrate}
\mathcal{Q}^{\rm U}={}&\frac{1}{1+3w^2}\Big[\left((3w^2+4w+1)\rho f_{,\rho\rho}+4w f_{,\rho}\right)\dot{\rho}\\
&\qquad\qquad+3H\rho f_{,\rho}(3w^2+4w+1)\Big].
\end{aligned}
\end{equation}
Here, $f_{,\rho}$ denotes the derivative with respect to $\rho$ and
$H=\dot{a}/a$ is the Hubble parameter. So far, a single conventional sector
$T_{\mu\nu(1)}$ has been considered; we now add a second sector together with
its modification partner. A sufficient choice that produces linear background
kernels on a fixed-sign density interval is $f$ linear in $\rho$, as in
scale-independent EMSG~\cite{Akarsu:2018aro}:
\begin{align}
f=&\alpha_1\sqrt{T_{\mu\nu(1)}T^{\mu\nu}_{(1)}}+\alpha_2\sqrt{T_{\mu\nu(2)}T^{\mu\nu}_{(2)}}\nonumber\\
=&\alpha_1\sqrt{1+3w_1^2}\lvert\rho_1\rvert+\alpha_2\sqrt{1+3w_2^2}\lvert\rho_2\rvert,\label{eq:scaleinde}
\end{align}
where $\lvert\rho_i\rvert=\rho_i$ if $\rho_i>0$ with $i=1,2$.
Accordingly, all equations derived below refer to the positive-density branch,
on which $f_{,\rho_i}=\alpha_i\sqrt{1+3w_i^2}={\rm const.}$ and
$f_{,\rho_i\rho_i}=0$; the couplings $\alpha_i$ are dimensionless. Here and
below, ``positive-density branch'' means $\rho_i>0$ for the
conventional sources; it does not require the algebraic EMSF contributions to
be positive. A continuation through $\rho_i=0$ is not covered because
$f$ is nondifferentiable there and $f_{,\rho}$ changes sign. To isolate the
consequences of the conservation assignment with the smallest parameter set,
we take a common coupling within the two-species EMSG sector,
$\alpha_1=\alpha_2\equiv\alpha$. This restriction is substantive: with
$\alpha_1\ne\alpha_2$, sector separation instead constrains a weighted sum of
the EMSF currents, and the diagonalized conventional kernels generally depend
on the ratio $\alpha_1/\alpha_2$.

Total conservation as per the temporal component of Eq.~\eqref{eq:bianchi} leads to
\begin{equation}
\begin{aligned}
\label{eq:wrons}
&\dot{\rho}_1+3H(w_1+1)\rho_1+\dot{\rho}_2+3H(w_2+1)\rho_2\\
&+\alpha\big[A_1\dot{\rho}_1+3H(w_1+1)B_1\rho_1+A_2\dot{\rho}_2 \\
&\quad\quad+3H(w_2+1)B_2\rho_2\big]=0.
\end{aligned}
\end{equation}
The EMSF contribution, represented by the terms proportional to $\alpha$, is
related to the projected current~\eqref{eq:creationrate} by
 \begin{align}
&\alpha\big[A_1\dot{\rho}_1+3H(w_1+1)B_1\rho_1
+A_2\dot{\rho}_2\nonumber\\
&\hspace{4.6em}+3H(w_2+1)B_2\rho_2\big]
=-\mathcal{Q}_{0(1)}^{\rm mod}-\mathcal{Q}_{0(2)}^{\rm mod},\label{eq:cont1}
\end{align}
where $A_i$ and $B_i$ are dimensionless species-dependent constants. For the
species with EoS parameter $w_1$,
\begin{align}
\label{eq:coef1}
A_{1}=&\frac{4w_{1}}{\sqrt{3w_{1}^2+1}}, \quad B_{1}=\frac{3w_{1}+1}{\sqrt{3w_{1}^2+1}}.
\end{align}
The species-2 expressions follow by interchanging $1\leftrightarrow2$. This
notation will be used below.

By combining Eq.~\eqref{eq:wrons} with Eq.~\eqref{eq:cont1}, we derive
\begin{equation}
\begin{aligned}
\dot{\rho}_1+3H(w_1+1)\rho_1
&+\dot{\rho}_2\\
&+3H(w_2+1)\rho_2=-\mathcal{Q}_{0(1)}-\mathcal{Q}_{0(2)}.
\end{aligned}
\label{eq:cont2}
\end{equation}
We now impose the sector-separation closure: the conventional and EMSF sectors
are conserved separately according to Eqs.~\eqref{eq:sectori} and
\eqref{eq:sectorii}. This is the sector-separated representative assignment of
Sect.~\ref{sec:iii}, not an on-shell identity of the separable action. Related
nonminimal exchange in scale-independent EMSG is discussed in
Refs.~\cite{Akarsu:2018aro,Akarsu:2023nyl}.

With Eq.~\eqref{eq:sectorii}, Eq.~\eqref{eq:cont1} has the form
$\alpha\mathcal F=0$. At $\alpha=0$ the EMSF sector vanishes and this equation
is an identity. The closure then supplies only conservation of the sum of the
conventional stresses; separate conservation of the two uncoupled GR species
follows only after reinstating their matter equations from the minimal
$\alpha=0$ action. For $\alpha\ne0$, division by $\alpha$ instead imposes
\begin{equation}
\begin{aligned}
A_1\dot{\rho}_1+3H(w_1+1)B_1\rho_1
&+A_2\dot{\rho}_2\\
&+3H(w_2+1)B_2\rho_2=0. 
\end{aligned}
\label{eq:cont11}
\end{equation}
Equation~\eqref{eq:cont11} is therefore a constraint only on the
nonzero-coupling closure family. Its survival after division by $\alpha$, and
its disappearance when $\alpha=0$ is imposed first, mean that this
conservation prescription loses one constraint at $\alpha=0$; this is
unrelated to the rank loss of the density-reconstruction matrix in
Sect.~\ref{sec:alphaparticular}. This is the sense in which the closure
is singular at the GR point; it is not a singularity of the underlying EMSG
action. Whether a special solution nevertheless has a continuous uncoupled
GR limit must be checked separately.%

 Separate conservation of the conventional sector,
Eq.~\eqref{eq:sectori}, gives
\begin{equation}
\begin{aligned}
\label{eq:cont3} 
\dot{\rho}_1+3H(w_1+1)\rho_1
+\dot{\rho}_2+3H(w_2+1)\rho_2=0,
\end{aligned}
\end{equation}
as follows directly from Eq.~\eqref{eq:cont2}.

Solving Eqs.~\eqref{eq:cont11} and \eqref{eq:cont3} for
$\dot\rho_1$ and $\dot\rho_2$ gives the equivalent system
\begin{align}
\label{eq:Q1}
\dot{\rho}_1+3H(w_1+1)\rho_1&=H(\beta\rho_1+\zeta\rho_2),\\
\label{eq:Q2}
\dot{\rho}_2+3H(w_2+1)\rho_2&=-H(\beta\rho_1+\zeta\rho_2),
\end{align}
where the species-dependent constants $\beta(w_1,w_2)$ and
$\zeta(w_1,w_2)$ are
 \begin{equation}
\begin{aligned}
 \label{eq:f1g}
\beta(w_1,w_2)&=3(w_1+1)\frac{A_1(w_1)-B_1(w_1)}{A_1(w_1)-A_2(w_2)}\,,\\ 
\zeta(w_1,w_2)&=3(w_2+1)\frac{A_2(w_2)-B_2(w_2)}{A_1(w_1)-A_2(w_2)}.
\end{aligned}
\end{equation}
Equivalently, in terms of the source EoS parameters,
 \begin{equation}
\begin{aligned}
 \label{eq:eqf1g}
\beta(w_1,w_2)&=\frac{3(w_1^2-1)\sqrt{3w_2^2+1}}{4\left[w_1\sqrt{3w_2^2+1}-w_2\sqrt{3w_1^2+1}\right]}\,,\\ 
\zeta(w_1,w_2)&=\frac{3(w_2^2-1)\sqrt{3w_1^2+1}}{4\left[w_1\sqrt{3w_2^2+1}-w_2\sqrt{3w_1^2+1}\right]}.
\end{aligned}
\end{equation}
These expressions are valid for $A_1\neq A_2$, equivalently $w_1\neq w_2$. The equal-EoS case must be treated directly from the original sector-separated conservation equations and is not described by the coefficients $\beta$ and $\zeta$ given above.\footnote{For $w_1=w_2=w$, combining Eqs.~\eqref{eq:cont11} and~\eqref{eq:cont3} yields $H(1-w^2)(\rho_1+\rho_2)=0$. Hence, for $H\neq0$ and $\rho_1+\rho_2\neq0$, the nontrivial equal-EoS cases are restricted to $w=\pm1$.}
Equations~\eqref{eq:Q1} and \eqref{eq:Q2} have the same linear structure as
the phenomenological interaction model studied by Barrow and Clifton
\cite{Barrow:2006hia}. There the coefficients are arbitrary constants; here
$\zeta$ and $\beta$ become functions of the species EoS parameters after the
scale-independent EMSF prescription, common coupling, and sector-separation
closure have all been specified. They are therefore closure-dependent but
species-determined, not quantities fixed by the EMSG action alone. The coupling
$\alpha$ remains explicit in the gravitational equations through the EMSF
contributions.
An important consequence of the common-coupling and sector-separation
assumptions is that $\alpha$ cancels from the conventional-sector continuity
equations for every $\alpha\neq0$. Hence, $\beta$ and $\zeta$ are not
perturbatively small interaction coefficients controlled by $|\alpha|$.
Rather, $\alpha$ controls the explicit EMSF amplitudes in the gravitational
equations, while the effective exchange rates are fixed by the equations of
state within this closure family.
Accordingly, for nonvanishing closure-generated exchange, making $|\alpha|$
arbitrarily small does not suppress that exchange. The generic nonzero-$\alpha$
sector-separated family therefore fails to approach the uncoupled GR
conservation system. The latter follows by setting $\alpha=0$ before the
division and reinstating the matter equations of the uncoupled action.
There is an important exception: for finite $w_1\ne w_2$,
Eq.~\eqref{eq:eqf1g} gives $\beta=\zeta=0$ precisely for the unordered pair
$\{w_1,w_2\}=\{-1,1\}$. This stiff--vacuum pair has the standard separate
dilution laws and a continuous GR background limit, as shown in
Sect.~\ref{sec:csga1}. The loss of this constraint at $\alpha=0$ does not imply
discontinuity of every special background.%

\paragraph{Pressure mismatch and the origin of the kernel.}
\label{sec:pressure-mismatch}
A dimensionless measure of departure from the parent equation of state,
without dividing by the partner energy density, is
\begin{equation}
\delta\pi_i\equiv
\frac{p_i^{\rm EMSF}-w_i\rho_i^{\rm EMSF}}{\alpha\rho_i}
=\frac{1-w_i^2}{\sqrt{1+3w_i^2}}.
\label{eq:pressure-mismatch}
\end{equation}
Here $\rho_i^{\rm EMSF}=\alpha A_i\rho_i$ and
$p_i^{\rm EMSF}=\alpha\sqrt{1+3w_i^2}\rho_i$, as defined below in
Eq.~\eqref{eq:emsf-component-definitions}; $\alpha\ne0$ and $\rho_i>0$.
Unlike the partner's pressure-to-density ratio, this quantity remains
regular for dust. Equation~\eqref{eq:f1g} becomes
\begin{equation}
\beta=-\frac{3\delta\pi_1}{A_1-A_2},\qquad
\zeta=-\frac{3\delta\pi_2}{A_1-A_2},
\label{eq:kernel-mismatch-coefficients}
\end{equation}
so the conventional exchange is
\begin{equation}
Q\equiv H(\beta\rho_1+\zeta\rho_2)
=-\frac{3H\sum_i(p_i^{\rm EMSF}-w_i\rho_i^{\rm EMSF})}
{\alpha(A_1-A_2)}.
\label{eq:kernel-mismatch-current}
\end{equation}
Separate conservation of the two sector sums compensates the pressure
mismatch through exchange between the conventional species. The common
factor $\alpha$ appears both in that mismatch and in the difference
$\alpha(A_1-A_2)$ of the EMSF energy coefficients $\alpha A_i$, and cancels
when the equations are solved. Each coefficient vanishes precisely at
$w_i=\pm1$, where its partner follows the parent equation of state. For the
action-derived stresses of Eq.~\eqref{eq:action-stress-rescaling},
$T^{\rm mod}_{\mu\nu}\propto T_{\mu\nu}$ makes the mismatch vanish
identically, consistent with the absence of exchange found there for
$k_1\ne k_2$. For unequal couplings $\alpha_1\ne\alpha_2$, the
algebra leading to Eq.~\eqref{eq:kernel-mismatch-coefficients} gives
\begin{equation}
\beta=-\frac{3\alpha_1\delta\pi_1}{\alpha_1A_1-\alpha_2A_2},\qquad
\zeta=-\frac{3\alpha_2\delta\pi_2}{\alpha_1A_1-\alpha_2A_2},
\label{eq:unequal-coupling-kernel}
\end{equation}
provided $\alpha_1A_1-\alpha_2A_2\ne0$. The singular denominator requires
the original conservation equations; unequal $w_i$ alone does not exclude
it. This extension makes explicit the coupling-ratio dependence; all
subsequent solutions retain the common nonzero coupling.

The role of the conservation assignment can be isolated while holding the
same algebraic stress prescription fixed. The species-paired rule
\eqref{eq:minimal-onshell} would instead give, for each species,
\begin{equation}
(1+\alpha A_i)\dot\rho_i
+3H(1+w_i)(1+\alpha B_i)\rho_i=0.
\label{eq:paired-algebraic-conservation}
\end{equation}
For $1+\alpha A_i\ne0$, its conventional-sector source is
\begin{equation}
\dot\rho_i+3H(1+w_i)\rho_i
=\frac{3\alpha H(1+w_i)(A_i-B_i)}{1+\alpha A_i}\rho_i,
\label{eq:paired-algebraic-source}
\end{equation}
which tends to zero with $\alpha$ for finite regular background variables.
Thus the $\alpha$-independent exchange is not forced by the algebraic EMSF
stress itself. This is a comparison of conservation assignments at fixed
stress prescription, not an off-shell derivation of that prescription or a
microscopic completion of either background system.%

Table~\ref{tab:attribution} summarizes the four combinations,
distinguishing the effects of the stress prescription from those
of the conservation assignment.

\begin{table*}[t]
\caption{Attribution of background behavior for the specified
constant-equation-of-state constructions. Action-derived entries refer to
the explicit Brown or regular Schutz actions of
Eqs.~\eqref{eq:brown-explicit-action}--\eqref{eq:schutz-explicit-action},
not to all possible fluid actions. The species-paired rule is action selected
only in the first row; in the second it is a comparison at fixed algebraic
stress. The conditions $1+\alpha k_i\ne0$, the regular Schutz domain, and
the exception $k_1=k_2$ stated in the text are essential.}
\label{tab:attribution}
\begin{ruledtabular}
\begin{tabular}{p{0.19\textwidth}p{0.35\textwidth}p{0.35\textwidth}}
\raggedright Stress prescription & \raggedright Species-paired conservation & \raggedright Sector-separated closure, $\alpha\ne0$ \tabularnewline
\hline
\raggedright Explicit action-derived stress &
\raggedright GR dilution for $1+\alpha k_i\ne0$; gravitational weights $1+\alpha k_i$ &
\raggedright Zero exchange if $k_1\ne k_2$; undetermined by the closure alone if $k_1=k_2$ \tabularnewline[3pt]
\raggedright Algebraic stress, Eq.~\eqref{eq:fluid-hessian-prescription} &
\raggedright Modified dilution, Eq.~\eqref{eq:paired-algebraic-source}; regular uncoupled limit as $\alpha\to0$ &
\raggedright $\alpha$-independent Barrow--Clifton kernel; generic singular limit, with the zero-exchange stiff--vacuum exception
\end{tabular}
\end{ruledtabular}
\end{table*}

For compactness, define
\begin{equation}
K_i\equiv1+\alpha A_i,
\qquad
P_i\equiv w_i+\alpha\sqrt{1+3w_i^2}.
\label{eq:KP-definitions}
\end{equation}
The Friedmann and pressure equations are then
\begin{align}
\label{eq:0fr2}
3H^2&=\kappa(K_1\rho_1+K_2\rho_2),
\end{align}
\begin{align}
\label{eq:0frpres}
-2\dot{H}-3H^2&=\kappa(P_1\rho_1+P_2\rho_2),
\end{align}
satisfying the separated continuity equations given in
Eqs.~\eqref{eq:cont11}--\eqref{eq:cont3} [or equivalently
\eqref{eq:Q1}--\eqref{eq:Q2}], which hold for $\alpha\neq0$. For each species,
the algebraic EMSF coefficients in the two gravitational equations define
\begin{equation}
\rho_i^{\rm EMSF}\equiv\alpha A_i\rho_i,
\qquad
p_i^{\rm EMSF}\equiv\alpha\sqrt{1+3w_i^2}\,\rho_i.
\label{eq:emsf-component-definitions}
\end{equation}
For $\alpha w_i\ne0$, their componentwise ratio is
\begin{equation}
\label{eq:wemsf}
    w_i^{\rm EMSF}\equiv
    \frac{p_i^{\rm EMSF}}{\rho_i^{\rm EMSF}}
    =\frac{1+3w_i^2}{4w_i}.
\end{equation}
The EMSF partner of dust ($w=0$) instead has a vanishing contribution to the
effective energy density but a nonzero contribution to the effective pressure.
Its componentwise EoS ratio and density-dilution parameter are therefore
undefined rather than finite. Radiation ($w=1/3$) has a stiff-like EMSF
partner, while cosmic strings ($w=-1/3$) have a vacuum-like partner. Vacuum
energy ($w=-1$) and stiff matter ($w=1$) preserve their respective EoS ratios.

We next solve Eqs.~\eqref{eq:Q1}--\eqref{eq:Q2} as functions of the scale
factor. Eliminating one density in turn, as in Ref.~\cite{Barrow:2006hia}, gives
\begin{align}
\rho_1''+C(w_1,w_2)\rho_1'+2D(w_1,w_2)\rho_1&=0,
\label{eq:dr1}\\
\rho_2''+C(w_1,w_2)\rho_2'+2D(w_1,w_2)\rho_2&=0.
\label{eq:dr2}
\end{align}
Here $^\prime$ denotes differentiation with respect to $\ln a$, and
$C(w_1,w_2)$ and $D(w_1,w_2)$ are defined by
\begin{align}
\label{eq:C}
C(w_1,w_2)&=\zeta-\beta+3(w_1+w_2+2),\\
D(w_1,w_2)&=\frac{3}{2}\big[\zeta(w_1+1)-\beta(w_2+1)\nonumber\\
&\hspace{5.3em}+3(w_1+1)(w_2+1)\big].
\label{eq:D}
\end{align}
For $w_1\ne w_2$, the same coefficient can be factorized as
\begin{equation}
2D=9(1+w_1)(1+w_2)\frac{B_2-B_1}{A_2-A_1}.
\label{eq:D-factorized}
\end{equation}
Thus $D$ vanishes not only when either species is vacuum energy. For a finite
unequal-EoS pair, its complete zero set is
\begin{equation}
\begin{gathered}
w_1=-1,\quad\text{or}\quad w_2=-1,\\
\text{or}\quad 1+w_1+w_2-3w_1w_2=0,\\
w_1,w_2>\tfrac13\quad\text{on the last locus}.
\end{gathered}
\label{eq:D-zero-locus}
\end{equation}
The last condition is the unequal, nonvacuum solution of $B_1=B_2$,
parametrized by $w_2=(1+w_1)/(3w_1-1)$ with $w_1>\tfrac13$; for example, it
contains $(w_1,w_2)=(1/2,3)$. Whenever $D=0$ and $C\ne0$, the
characteristic exponents are $0$ and $C$, and hence
\begin{equation}
\rho_i(a)=c_{i0}+c_{i1}a^{-C},
\label{eq:D-zero-density}
\end{equation}
with amplitudes constrained by the original first-order system. In
particular, Eq.~\eqref{eq:cont3} forces $(1+w_1)c_{10}=-(1+w_2)c_{20}$: on the
nonvacuum locus, whenever the constant mode is present, its two density
amplitudes have opposite signs, while on the vacuum lines the nonvacuum species
has no constant part, consistent with the
explicit solutions of Sect.~\ref{sec:csga1}. Moreover, $C$ is strictly positive
on the entire zero set: $C=3+\tfrac32\sqrt{3w^2+1}\geq\tfrac92$ on the vacuum
lines, with $w$ the EoS parameter of the nonvacuum species. On the nonvacuum
locus, substituting $w_2=(1+w_1)/(3w_1-1)$ gives
\begin{equation}
C=\frac32(2+w_1+w_2)
=6+\frac{9(w_1-1)^2}{2(3w_1-1)}>6,
\label{eq:D-zero-C-bound}
\end{equation}
where the strict inequality follows from $w_1>1/3$ and the exclusion of the
equal-EoS point $w_1=w_2=1$. The logarithmic solution that would replace
$a^{-C}$ at $C=0$ therefore never arises on the zero set, and the $a^{-C}$ mode
always decays. On a branch for which both conventional densities are positive
at $a=1$, a nonzero constant mode on the nonvacuum locus makes the formal
density solution for one of them vanish at a finite scale factor $a_*>1$.
This zero is reached by the cosmological solution only if the expanding FLRW
branch remains in the real domain $H^2>0$ up to $a_*$. Otherwise the branch
encounters $H^2=0$ and terminates or turns around first. In the generic case
the energy densities are
\begin{align}
\label{eq:rr1}
&\rho_1(a)=d_+a^{-\gamma_+}+d_-a^{-\gamma_-},\\
\label{eq:rr2}
&\rho_2(a)=f_+a^{-\gamma_+}+f_-a^{-\gamma_-},
\end{align}
where both densities share the exponents $\gamma_\pm$. Their initial mode
amplitudes $d_\pm$ and $f_\pm$ are not independent: they are constrained mode
by mode by the original first-order system. The exponents are
\begin{equation}
\begin{aligned}
\gamma_{\pm}(w_1,w_2)=\frac{1}{2}
\left(C\pm\sqrt{C^2-8D}\right).
\end{aligned}
\label{eq:rr4}
\end{equation} 

For finite real $w_1\ne w_2$, these exponents are always real and distinct.
Indeed, substituting Eq.~\eqref{eq:eqf1g} into
Eqs.~\eqref{eq:C}--\eqref{eq:D} gives the identity
\begin{equation}
\begin{aligned}
C^2-8D&=(C-6)^2\\
&\quad+9\sqrt{(1+3w_1^2)(1+3w_2^2)}>9.
\end{aligned}
\label{eq:char-discriminant-identity}
\end{equation}
The strict inequality follows because an unequal pair cannot have
$w_1=w_2=0$, so the square-root factor exceeds unity. In particular,
$\gamma_+-\gamma_->3$ throughout this domain.%

Substituting the superposed solutions~\eqref{eq:rr1} and \eqref{eq:rr2}, with the exponents \eqref{eq:rr4}, into Eqs.~\eqref{eq:Q1} and \eqref{eq:Q2} and equating the coefficients of the independent modes leads to the relation
\begin{align}
    d_{\pm}=\frac{\eta_{\pm}}{\beta}f_{\pm},
\end{align}
with
\begin{equation}
\begin{aligned}
\eta_{\pm}(w_1,w_2)={}&\gamma_{\pm}(w_1,w_2)
-3(w_2+1)-\zeta.
\end{aligned}
\label{eq:qpm}
\end{equation}
This representation requires $\beta\neq0$; the inequality $\eta_+\ne \eta_-$ is
automatic in the finite unequal-EoS domain. Cases with $\beta=0$
(i.e., $w_1=\pm1$) must be treated directly from
Eqs.~\eqref{eq:Q1}--\eqref{eq:Q2}.%

Incorporating the constrained amplitudes into the Friedmann
equation~\eqref{eq:0fr2} gives
\begin{align}
3H^2={}&\kappa K_1
\left(d_{+}a^{-\gamma_{+}}+d_{-}a^{-\gamma_{-}}\right)\nonumber\\
&+\kappa K_2
\left(f_{+}a^{-\gamma_{+}}+f_{-}a^{-\gamma_{-}}\right),
    \label{eq:gfdet}
\end{align}
with present-day values $\rho_{10}=d_++d_-$ and
$\rho_{20}=f_++f_-$. We normalize $a(t_0)=1$, where $t_0$ is the present
time, and define $H_0=H(t_0)$, $\rho_{\rm c0}=3H_0^2/\kappa$, and
$\Omega_{i0}=\rho_{i0}/\rho_{\rm c0}$. The present-day density parameters are
\begin{align}
    &\Omega_1(w_1,w_2;1)=\frac{\kappa}{3H_0^2}\left[d_++d_-\right]=\Omega_{10},\label{eq:edpt1}\\
    &\Omega_2(w_1,w_2;1)=\frac{\kappa}{3H_0^2}\left[f_++f_-\right]=\Omega_{20}.\label{eq:edpt2}
\end{align}
The corresponding EMSF-partner density parameters are
\begin{align}
\Omega_1^{\rm EMSF}(\alpha,w_1,w_2;1)
&=\alpha A_1\Omega_{10}
=\Omega_{10}^{\rm EMSF}(\alpha,w_1),\label{eq:edpt3}\\
\Omega_2^{\rm EMSF}(\alpha,w_1,w_2;1)
&=\alpha A_2\Omega_{20}
=\Omega_{20}^{\rm EMSF}(\alpha,w_2).
    \label{eq:edpt4}    
\end{align}
For compactness, define the present-day conventional-density ratio
\begin{equation}
\mathcal R_{12}\equiv
\frac{\Omega_{20}}{\Omega_{10}}
=\frac{1-(1+\alpha A_1)\Omega_{10}}
{(1+\alpha A_2)\Omega_{10}}.
\label{eq:R12}
\end{equation}
The second equality uses spatial flatness and assumes
$(1+\alpha A_2)\Omega_{10}\ne0$. A dust-safe generic parametrization is
$(\alpha,\mathcal R_{12})$, for which
\begin{equation}
\Omega_{10}=\frac{1}{K_1+K_2\mathcal R_{12}},\qquad
\Omega_{20}=\frac{\mathcal R_{12}}{K_1+K_2\mathcal R_{12}}.
\label{eq:R12-normalization}
\end{equation}
The normalized conventional densities are then
\begin{align}
U_1(a)={}&\frac{\eta_+a^{-\gamma_+}-\eta_-a^{-\gamma_-}}
{\eta_+-\eta_-}\nonumber\\
&-\frac{\eta_+\eta_-\mathcal R_{12}}
{\beta(\eta_+-\eta_-)}
\left(a^{-\gamma_+}-a^{-\gamma_-}\right),\label{eq:U1compact}\\
U_2(a)={}&\frac{\eta_+a^{-\gamma_-}-\eta_-a^{-\gamma_+}}
{\eta_+-\eta_-}\nonumber\\
&-\frac{\beta\mathcal R_{12}^{-1}}{\eta_+-\eta_-}
\left(a^{-\gamma_-}-a^{-\gamma_+}\right).
\label{eq:U2compact}
\end{align}
Here, by construction, $U_1(1)=U_2(1)=1$ at the present epoch. This compact
representation assumes $\beta(\eta_+-\eta_-)\mathcal R_{12}\ne0$ and all explicit
denominators in Eqs.~\eqref{eq:R12}--\eqref{eq:R12-normalization} to be
nonzero; singular cases must be obtained directly from the first-order system.
These expressions are local background solutions. Their admissible interval is
the connected range of $a$ containing $a=1$ on which
$\rho_1(a)>0$, $\rho_2(a)>0$, and $H^2(a)>0$. Nonnegative present-day
density parameters alone do not guarantee these inequalities at all times.
Rearranging Eq.~\eqref{eq:gfdet} with
Eqs.~\eqref{eq:edpt1}--\eqref{eq:edpt4}, the Friedmann equation, which
distinguishes the source and EMSF-partner contributions to the expansion,
becomes
\begin{equation}
\begin{aligned}
\frac{H^2}{H_0^2}={}&
(\Omega_{10}+\Omega_{10}^{\rm EMSF})U_1(a)\\
&+(\Omega_{20}+\Omega_{20}^{\rm EMSF})U_2(a),
\end{aligned}
\label{eq:gfdet2}
\end{equation}
with the present-day consistency relation ($a = 1$):
\begin{equation}
\begin{aligned}
1={}&\Omega_{10}+\Omega_{10}^{\rm EMSF}(\alpha,w_1)\\
&+\Omega_{20}+\Omega_{20}^{\rm EMSF}(\alpha,w_2).
\end{aligned}
\end{equation}
The background is therefore encoded by two normalized functions, $U_1(a)$ and
$U_2(a)$. Although the bookkeeping contains two conventional fluids and two
EMSF partners, the scale-factor dependence has only the two common modes
$a^{-\gamma_-}$ and $a^{-\gamma_+}$. The EMSF partners introduce no
independent background degrees of freedom: their stresses are algebraic in the
conventional densities, whose evolution is governed by the two first-order
equations~\eqref{eq:intsysi} and \eqref{eq:intsysii}.

\section{Cosmological solutions}
\label{sec:cossol}
We first consider pairs containing vacuum energy or a stiff fluid, for which the
interaction coefficients and exact solutions simplify, and then return to a
general $(w_1,w_2)$ pair. The effective EoS parameters introduced for the
coupled modes characterize the dilution of the conventional densities; they
should not be confused with the EoS ratios of the EMSF partners in
Eq.~\eqref{eq:wemsf}.

\subsection{\texorpdfstring{$w$}{w}-fluid--vacuum-energy pair}
\label{sec:csga1}

We consider a conventional source with a general EoS parameter
$w_1=w={\rm const.}$ (excluding vacuum energy, $w=-1$), and conventional
vacuum energy $w_2=-1$. For $w\ne0$, their respective EMSF-partner EoS ratios
are
\begin{align}
w_1=w,&\qquad w_1^{\rm EMSF}=\frac{1+3w^2}{4w},\nonumber\\
w_2&=w_2^{\rm EMSF}=-1.
\end{align}
For $w=0$, the dust exception below Eq.~\eqref{eq:wemsf} applies.
For a $w$-fluid--vacuum pair, Eq.~\eqref{eq:eqf1g} gives
$\zeta_{\rm v}(w)=0$ and hence $D_{\rm v}(w)=0$ by Eq.~\eqref{eq:D}.
Equations~\eqref{eq:Q1} and \eqref{eq:Q2} reduce to
\begin{align}
\label{eq:rhomod}
\dot{\rho}_w+3H(w+1)\rho_w= &\,\beta_{\rm v}(w) H\rho_w,\\
\dot{\rho}_{\rm v}=&-\beta_{\rm v}(w) H\rho_w,
\end{align}
where
\begin{equation}
\beta_{\rm v}(w)=3w-\frac{3}{2}\sqrt{3w^2+1}.
\label{eq:weffdef}
\end{equation}
Define the effective dilution parameter
\begin{equation}
\label{eq:weffw}
w_{\rm v,eff}(w)=\frac{1}{2}\sqrt{3w^2+1},
\end{equation}
and the abbreviations
\begin{align}
n_{\rm v}&\equiv3\left[1+w_{\rm v,eff}(w)\right],\nonumber\\
c_{\rm v}&\equiv
\frac{w_{\rm v,eff}(w)-w}{1+w_{\rm v,eff}(w)}.
\label{eq:vacuum-domain-definitions}
\end{align}
The energy densities are
\begin{align}
\rho_w(a)&=\rho_{w0}a^{-n_{\rm v}},
\label{eq:rhow}\\
\rho_{\rm v}(a)&=\rho_{\rm v0}
+c_{\rm v}\rho_{w0}\left(1-a^{-n_{\rm v}}\right).
\label{eq:beta}
\end{align}
The domain requires special care. For $-1<w<1$, one has $c_{\rm v}>0$.
For positive present-day densities, the vacuum density reaches zero at
\begin{equation}
a_{{\rm v},*}=\left(1+
\frac{\rho_{\rm v0}}{c_{\rm v}\rho_{w0}}
\right)^{-1/n_{\rm v}}<1.
\label{eq:vacuum-branch-endpoint}
\end{equation}
The positive-density solution is therefore valid only for
$a>a_{{\rm v},*}$. A continuation to earlier times requires a separate
piecewise derivation across the nondifferentiable point of $f\propto
|\rho_{\rm v}|$.
For $w\ne0$, Eq.~\eqref{eq:emsf-component-definitions} shows that
$\rho_w^{\rm EMSF}=\alpha A(w)\rho_w$ is a constant multiple of $\rho_w$ and
therefore has the same scale-factor dependence. Purely as a dilution label,
\begin{equation}
\label{eq:wefff}
w_{\rm v,eff}^{\rm EMSF}(w)=w_{\rm v,eff}(w)
=\frac{1}{2}\sqrt{3w^2+1}\qquad(w\ne0).
\end{equation}
For dust, however, $\rho_{\rm CDM}^{\rm EMSF}=0$ identically while
$p_{\rm CDM}^{\rm EMSF}=\alpha\rho_{\rm CDM}$; hence no EMSF-partner dilution
parameter is defined.

To display the Friedmann equation compactly, define
\begin{align}
\mathcal A_{\rm v}\equiv{}&
\frac{w-w_{\rm v,eff}}{1+w_{\rm v,eff}}
\frac{\Omega_{w0}}{\Omega_{{\rm v}0}}
=-c_{\rm v}\frac{\Omega_{w0}}{\Omega_{{\rm v}0}},\nonumber\\
\mathcal J_{\rm v}(a)\equiv{}&
\mathcal A_{\rm v}\left(1-a^{-n_{\rm v}}\right).
\label{eq:Jv}
\end{align}
Then Eq.~\eqref{eq:gfdet2} becomes
\begin{equation}
\begin{aligned}
\frac{H^2}{H_0^2}={}&
(\Omega_{w0}+\Omega_{w0}^{\rm EMSF})a^{-n_{\rm v}}\\
&+(\Omega_{\rm v0}+\Omega_{\rm v0}^{\rm EMSF})
\left[1-\mathcal J_{\rm v}(a)\right].
\end{aligned}
\label{eq:wvfr1}
\end{equation}
Here $\Omega_{{\rm v}0}\ne0$ is assumed in the compact definition; a vanishing
vacuum amplitude is handled directly from Eq.~\eqref{eq:gfdet2}. The energy
density parameters are defined as in
Eqs.~\eqref{eq:edpt1}--\eqref{eq:edpt4} and fulfill
\begin{equation}
\begin{aligned}
        1=&\Omega_{w0}+\Omega_{w0}^{\rm EMSF}+\Omega_{\rm v0}+\Omega_{\rm v0}^{\rm EMSF}.
\label{eq:fv}
\end{aligned}
\end{equation}

Writing Eq.~\eqref{eq:wvfr1} as
\begin{equation}
H^2=\chi_-+\chi_+a^{-n_{\rm v}},
\label{eq:wv-first-integral}
\end{equation}
the two coefficients are
\begin{equation}
\begin{aligned}
\chi_+={}&H_0^2\big[\Omega_{w0}+\Omega_{w0}^{\rm EMSF}
+\mathcal A_{\rm v}(\Omega_{\rm v0}+\Omega_{\rm v0}^{\rm EMSF})\big],\\
\chi_-={}&H_0^2(1-\mathcal A_{\rm v})
(\Omega_{\rm v0}+\Omega_{\rm v0}^{\rm EMSF}).
\end{aligned}
\end{equation}
For $\chi_+>0$, $\chi_->0$, and $n_{\rm v}\neq0$, the solution on any
connected interval not containing the integration point $t_{\rm B}$ is
\begin{equation}
a(t)=
\left(\frac{\chi_+}{\chi_-}\right)^{1/n_{\rm v}}
\left|
\sinh\left[
\frac{n_{\rm v}}{2}\sqrt{\chi_-}\,(t-t_{\rm B})
\right]
\right|^{2/n_{\rm v}}.
\label{eq:wv-hyperbolic-a}
\end{equation}
The normalization $a(t_0)=1$ fixes
\begin{equation}
\sinh^2\left[
\frac{n_{\rm v}}{2}\sqrt{\chi_-}\,(t_0-t_{\rm B})
\right]
=\frac{\chi_-}{\chi_+}.
\label{eq:wv-hyperbolic-normalization}
\end{equation}
The corresponding Hubble parameter is
\begin{equation}
H(t)=\sqrt{\chi_-}\,
\coth\left[
\frac{n_{\rm v}}{2}\sqrt{\chi_-}\,(t-t_{\rm B})
\right].
\label{eq:hwv1} 
\end{equation}
The intervals with $n_{\rm v}(t-t_{\rm B})>0$ and
$n_{\rm v}(t-t_{\rm B})<0$ describe the expanding and contracting branches,
respectively; they are related by time reflection about $t=t_{\rm B}$.
These formulae do not cover vanishing or negative $\chi_\pm$. The first
integral~\eqref{eq:wv-first-integral} nevertheless gives a useful formal
classification. The case $\chi_->0$, $\chi_+<0$ is cosh-type and has a nonzero
minimum scale factor, whereas $\chi_-<0$, $\chi_+>0$ is trigonometric and has a
finite maximum scale factor followed by recollapse. The limits
$\chi_-=0$, $\chi_+>0$ and $\chi_+=0$, $\chi_->0$ give power-law and de Sitter
evolution, respectively; $\chi_-<0$, $\chi_+<0$ gives no real FLRW interval.
These sign statements are kinematical. A physical fluid realization must also
satisfy $\rho_w>0$ and $\rho_{\rm v}>0$, including the endpoint restriction in
Eq.~\eqref{eq:vacuum-branch-endpoint}.

For a dust--vacuum pair, $w_{\rm v,eff}=1/2$ and
$\beta_{\rm v}=-3/2$, so that $\rho_{\rm CDM}\propto a^{-9/2}$ and the
closure produces $Q=-3H\rho_{\rm CDM}/2$ in the convention of
Eq.~\eqref{eq:rhomod}. This is a strongly nonstandard matter history. Bounds on
linear interacting-vacuum models depend on the kernel, normalization, sign
convention, momentum-transfer prescription, and data combination
\cite{Wands:2012vg,Kaeonikhom:2022ahf}; there is no model-independent preferred coupling that
can be transferred to the present construction. We therefore treat this result
as a formal illustration of the selected closure rather than as a viable
dark-sector model. The scaling holds for every nonzero $\alpha$, because
$\alpha$ cancels from the closure equations; it does not approach the
uncoupled $a^{-3}$ law as $\alpha\to0$ within this solution family.

For the stiff--vacuum pair one has
$\beta_{\rm v}=\zeta_{\rm v}=0$. Hence the diagonalized conventional
continuity equations are separately conserved at background order: there is no
energy exchange between the two conventional sources. The stiff component
obeys the standard dilution law $\rho_{\rm s}\propto a^{-6}$
\cite{Zeldovich:1961sbr,Chavanis:2014lra}, while the vacuum density remains
constant. Their gravitational amplitudes are nevertheless rescaled by the
explicit EMSF contributions. %
More explicitly, at fixed positive conventional-density amplitudes,
\begin{equation}
3H^2=\kappa\left[(1+2\alpha)\rho_{{\rm s}0}a^{-6}
+(1-2\alpha)\rho_{\rm v0}\right].
\label{eq:stiff-vacuum-GR-limit}
\end{equation}
On a regular expanding interval near $\alpha=0$, this tends continuously to
the uncoupled stiff--vacuum GR background. It is the zero-exchange exception
to the generic nonvanishing-exchange limit; the constraint contained in
$\alpha\mathcal F=0$ is still lost at $\alpha=0$.%

\subsubsection{Vacuum--radiation energy exchange: an illustrative early-time scenario}
\label{sec:signchangeDE}

As a representative example, we consider a fluid pair consisting of radiation
($w=1/3$) and vacuum energy ($w=-1$), an interaction with a long history in
vacuum-decay cosmology \cite{Freese:1986dd,Barrow:2006hia}. Here
$\rho_{\rm r}$ denotes an interacting radiation-like component distinct from
the conventional photon bath $\rho_\gamma$.
Conventional baryons, cold dark matter (CDM), and photons are taken here to be minimally coupled,
without EMSF partners, and separately conserved. Only the
$(\rho_{\rm r},\rho_{\rm v})$ pair is assigned the common EMSF coupling. This
is therefore a hybrid extension of the two-species case study, not a universal
coupling of all material species. Within the selected sector-separation closure
and metric-variation prescription, the interacting radiation-like mode and its
EMSF partner share the dilution parameter
$w_{\rm v,eff}=w_{\rm v,eff}^{\rm EMSF}=1/\sqrt{3}$ from
Eq.~\eqref{eq:wefff}. The transfer coefficient is
$\beta_{\rm v}=1-\sqrt{3}\simeq-0.73$ and is negative throughout
$-1<w<1$. Thus energy flows from the interacting radiation-like component to
the vacuum sector, and $\rho_{\rm r}$ dilutes faster than its noninteracting
counterpart.

For convenience, we define $X(z)\equiv(1+z)^{3+\sqrt{3}}$. The corresponding energy densities
obtained from Eqs.~\eqref{eq:rhow} and \eqref{eq:beta} are
\begin{align}
\label{eq:rhow1}
\rho_{\rm r}(z) &= \rho_{\rm r0}X(z),\\
\label{eq:rhoV1}
\rho_{\rm v}(z) &= \rho_{\rm v0}
+\rho_{\rm r0}\left(\frac{2}{\sqrt{3}}-1\right)
\left[1-X(z)\right].
\end{align}

The corresponding Friedmann equation is
\begin{align}
\frac{H^2}{H_0^2}={}&
\left(\Omega_{\rm v0}+\Omega_{\rm v0}^{\rm EMSF}\right)
\nonumber\\
&\times
\left[1+\left(\frac{2}{\sqrt{3}}-1\right)
\frac{\Omega_{\rm r0}}{\Omega_{\rm v0}}\left(1-X(z)\right)\right]
\nonumber\\
&+\left(\Omega_{\rm r0}+\Omega_{\rm r0}^{\rm EMSF}\right)X(z)
\nonumber\\
&+\Omega_{\rm m0}(1+z)^3+\Omega_{\gamma0}(1+z)^4 .
\label{eq:vacrad-friedmann}
\end{align}
The present-day flatness condition is
\begin{equation}
1=\Omega_{\rm v0}+\Omega_{\rm v0}^{\rm EMSF}
+\Omega_{\rm r0}+\Omega_{\rm r0}^{\rm EMSF}
+\Omega_{\rm m0}+\Omega_{\gamma0}.
\label{eq:vacrad-flatness}
\end{equation}

\begin{figure*}[t!]
\includegraphics[width=\textwidth]{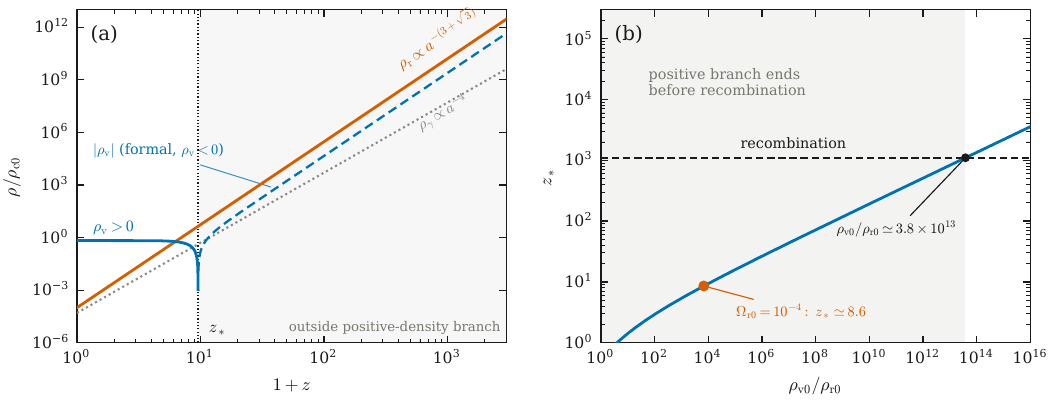}
\caption{Vacuum--radiation energy exchange of Sect.~\ref{sec:signchangeDE}.
\textbf{a}~The interacting component $\rho_{\rm r}\propto a^{-(3+\sqrt{3})}$ and
vacuum density $\rho_{\rm v}$ for $\Omega_{\rm r0}=10^{-4}$,
$\Omega_{\rm v0}=0.68$, and $\Omega_{\gamma0}=5\times10^{-5}$; the separately
conserved photon bath is dotted. The vacuum reaches
zero at $z_*$; the dashed $|\rho_{\rm v}|$ segment is a formal continuation
outside the positive-density branch.
\textbf{b}~The terminal redshift $z_*$ as a function of
$\rho_{\rm v0}/\rho_{\rm r0}$.
Reaching the benchmark $z_{\rm rec}=1100$ requires
$\rho_{\rm v0}/\rho_{\rm r0}\gtrsim3.8\times10^{13}$;%
 in the shaded region
the branch terminates earlier}
\label{fig:vacrad}
\end{figure*}

The interacting component dilutes as
$a^{-(3+\sqrt{3})}\simeq a^{-4.73}$. Figure~\ref{fig:vacrad}(a) shows the
corresponding evolution of $\rho_{\rm r}$ and $\rho_{\rm v}$.
Equation~\eqref{eq:rhoV1} shows that, for any $\rho_{\rm r0}>0$, the positive branch reaches
$\rho_{\rm v}=0$ at the finite redshift
\begin{equation}
(1+z_*)^{3+\sqrt{3}}=1+
\frac{\sqrt{3}}{2-\sqrt{3}}\frac{\rho_{\rm v0}}{\rho_{\rm r0}}.
\end{equation}
Continuing beyond this point requires a separate piecewise derivation from
$\sqrt{\rho_{\rm v}^2}=|\rho_{\rm v}|$.
Reaching the benchmark $z_{\rm rec}=1100$ on the positive branch requires%
 $\rho_{\rm v0}/\rho_{\rm r0}\gtrsim3.8\times10^{13}$;
see Fig.~\ref{fig:vacrad}(b).
For an order-unity present-day conventional vacuum density parameter
and order-unity gravitational weights, with the usual matter and photon
normalization, this bound makes the interacting radiation-like component
negligible compared with those matter and photon contributions at that
epoch. The density-ratio bound alone does not establish negligibility for
arbitrary gravitational weights and present-day normalizations.
Accordingly, no conclusion concerning
the $H_0$ tension or consistency with reconstructed interaction kernels follows
from this background solution. The example instead tests the domain of the
positive-density branch and shows that an early-time application would require
a piecewise continuation and a perturbative analysis.

\subsection{\texorpdfstring{$w$}{w}-fluid--stiff-fluid pair}
\label{sec:wstiff}
We consider a conventional source with a general EoS parameter
$w_1=w={\rm const.}$ (excluding stiff matter, $w=1$) and a stiff fluid
($w_2=1$). For $w\ne0$, their respective EMSF-partner EoS ratios are
\begin{align}
w_1=w,&\qquad w_1^{\rm EMSF}=\frac{1+3w^2}{4w},\nonumber\\
w_2&=w_2^{\rm EMSF}=1.
\end{align}
For $w=0$, the dust exception below Eq.~\eqref{eq:wemsf} applies.
For a $w$-fluid--stiff-fluid pair, Eq.~\eqref{eq:eqf1g} gives
$\zeta_{\rm s}(w)=0$. The integrated density solutions are
\begin{align}
\rho_w=&\rho_{w0}a^{-3(w+1)+\beta_{\rm s}(w)},\\
    \rho_{\rm s}=&\rho_{\rm s0}a^{-6}+\frac{\beta_{\rm s}(w)}{3(1-w)+\beta_{\rm s}(w)}\rho_{w0}\nonumber \\
    &\quad \quad\quad\quad\times\left(a^{-6}-a^{-\frac{3}{2}\left(2-\sqrt{3w^2+1}\right)}\right),
    \label{eq:sged}
    \end{align}
where
\begin{align}
\beta_{\rm s}(w)=\frac{3}{2}\left(2w+\sqrt{3w^2+1}\right).
\end{align}
For $w\ne0$, $\rho_w^{\rm EMSF}$ is a constant multiple of $\rho_w$ and has
the same scale-factor dependence. Purely as a dilution label,
\begin{equation}
w_{\rm s,eff}^{\rm EMSF}(w)=w_{\rm s,eff}(w)
=-\frac{1}{2}\sqrt{3w^2+1}\qquad(w\ne0).
\label{eq:weffs}
\end{equation}
For dust, $\rho_{\rm CDM}^{\rm EMSF}=0$ and the EMSF-partner dilution label is
undefined.
Define
\begin{align}
n_{\rm s}&\equiv3\left[1+w_{\rm s,eff}(w)\right]
=\frac{3}{2}\left(2-\sqrt{1+3w^2}\right),\nonumber\\
c_{\rm s}&\equiv
\frac{\beta_{\rm s}(w)}{3(1-w)+\beta_{\rm s}(w)}.
\label{eq:stiff-domain-definitions}
\end{align}
Equation~\eqref{eq:sged} can then be written as
\begin{equation}
\rho_{\rm s}(a)=a^{-6}\left[\rho_{\rm s0}
+c_{\rm s}\rho_{w0}\left(1-a^{6-n_{\rm s}}\right)\right].
\end{equation}
For $-1<w<1$, one has $c_{\rm s}>0$, and the stiff density reaches
zero at the finite future scale factor
\begin{equation}
a_{{\rm s},*}=\left(1+
\frac{\rho_{\rm s0}}{c_{\rm s}\rho_{w0}}
\right)^{1/(6-n_{\rm s})}>1.
\label{eq:stiff-branch-endpoint}
\end{equation}
Thus the positive-density solution is valid only for
$a<a_{{\rm s},*}$ unless a separate continuation through
$\rho_{\rm s}=0$ is supplied.
Define
\begin{align}
\mathcal J_{\rm s}(a)\equiv{}&
\frac{w-w_{\rm s,eff}}{1-w_{\rm s,eff}}
\frac{\Omega_{w0}}{\Omega_{{\rm s}0}}
\left(1-a^{6-n_{\rm s}}\right).
\label{eq:Js}
\end{align}
Equivalently,
$\mathcal J_{\rm s}=c_{\rm s}(\Omega_{w0}/\Omega_{{\rm s}0})
(1-a^{6-n_{\rm s}})$. This regular form assumes
$\Omega_{{\rm s}0}\ne0$; the zero-amplitude case follows directly from the
 density solution~\eqref{eq:sged}.
The Friedmann equation~\eqref{eq:gfdet2} is then
\begin{equation}
\begin{aligned}
\frac{H^2}{H_0^2}={}&
(\Omega_{w0}+\Omega_{w0}^{\rm EMSF})a^{-n_{\rm s}}\\
&+(\Omega_{\rm s0}+\Omega_{\rm s0}^{\rm EMSF})a^{-6}
\left[1+\mathcal J_{\rm s}(a)\right].
\end{aligned}
\label{eq:gwsfh}
\end{equation}
Here the energy density parameters are defined as in
Eqs.~\eqref{eq:edpt1}--\eqref{eq:edpt4} and fulfill
\begin{equation}
\begin{aligned}
        1=&\Omega_{w0}+\Omega_{w0}^{\rm EMSF}+\Omega_{\rm s0}+\Omega_{\rm s0}^{\rm EMSF}.
\label{eq:fvstiff}
\end{aligned}
\end{equation}
For example, the interacting dust and radiation-like modes dilute as
$a^{-3/2}$ and $a^{-3+\sqrt3}$, respectively, more slowly than their
noninteracting counterparts. For the vacuum--stiff pair,
$\beta_{\rm s}=0$: the conventional exchange kernel vanishes, although the
EMSF stress still modifies the gravitational amplitudes.

For $-1<w<1$, one has $\beta_{\rm s}>0$, so the conventional kernel transfers
energy from the stiff component to the $w$ fluid. For $w<-1$,
$\beta_{\rm s}<0$ and the direction is reversed. These statements refer only
to the positive-density interval identified above; they do not define a
covariant perturbative transfer model.

\subsection{\texorpdfstring{$(w_1,w_2)$}{(w1,w2)} fluid pair}
\label{sec:nvp}
For a general unequal-EoS pair, both scale-factor exponents in
Eq.~\eqref{eq:gfdet2} are nonzero precisely when $D\ne0$. On the additional
nonvacuum locus in Eq.~\eqref{eq:D-zero-locus}, one mode is constant and must be
handled by the corresponding limit. Away from that locus, define
$\Delta\gamma\equiv\gamma_+-\gamma_-$. On a connected expanding interval on
which
\begin{equation}
H^2(a)=\chi_-a^{-\gamma_-}+\chi_+a^{-\gamma_+}>0,
\label{eq:two-mode-H2}
\end{equation}
the real cosmic-time relation is unambiguously defined by
\begin{equation}
t-t_0=
\int_{1}^{a}
\frac{\mathrm{d}\widetilde a}
{\widetilde a\sqrt{
\chi_-\widetilde a^{-\gamma_-}
+\chi_+\widetilde a^{-\gamma_+}}}.
\label{eq:ggquadrature}
\end{equation}
When $\chi_+>0$, this quadrature admits the following principal-real-branch
representation in terms of the Gauss hypergeometric function:
\begin{align}
t-t_0&=\mathcal T(a)-\mathcal T(1),
\label{eq:gghs}
\\
\mathcal T(a)&\equiv
{}_2F_1\left(
1,1+\frac{\gamma_-}{2\Delta\gamma};
1+\frac{\gamma_+}{2\Delta\gamma};
-\frac{a^{\Delta\gamma}\chi_-}{\chi_+}
\right)\nonumber\\
&\quad\times
\frac{2\sqrt{\chi_-a^{-\gamma_-}+\chi_+a^{-\gamma_+}}}
{\gamma_+\chi_+a^{-\gamma_+}}.
\end{align}
The $\chi_{\pm}$ coefficients are composite, combining all coefficients
relevant to the scale factor for the case at hand:
\begin{equation}
\begin{aligned}
\chi_\pm={}&\frac{H_0^2}{\eta_+-\eta_-}
\left[\frac{\eta_\pm}{\beta}\left(1+\alpha A_1\right)
+\left(1+\alpha A_2\right)\right]\\
&\times\left(\pm\beta\Omega_{10}\mp \eta_\mp\Omega_{20}\right).
\end{aligned}
\label{eq:chipm}
\end{equation}
Here, $A_i=A_i(w_i)$, $\beta=\beta(w_1,w_2)$, and
$\eta_\pm=\eta_\pm(w_1,w_2)$ are defined in Eqs.~\eqref{eq:coef1},
\eqref{eq:eqf1g}, and \eqref{eq:qpm}, respectively. Hence, $\chi_\pm$ are
explicit functions of $(w_1,w_2,\alpha,\Omega_{10},\Omega_{20})$.
Equation~\eqref{eq:gghs} requires
$\beta(\eta_+-\eta_-)\gamma_+\chi_+\neq0$, as well as
\begin{equation}
1+\frac{\gamma_+}{2\Delta\gamma}
\notin\{0,-1,-2,\ldots\},
\end{equation}
unless the corresponding limiting form is taken. If $\chi_+<0$, the argument
of the Gauss function can lie on its principal branch cut even while $H^2>0$.
In that case Eq.~\eqref{eq:ggquadrature}, or an explicitly selected real
analytic continuation of Eq.~\eqref{eq:gghs}, must be used.
Vanishing denominators or a zero of $H^2$ likewise require the
appropriate limit or a separate branch analysis.

\paragraph{A fixed-width positive-density domain.}
\label{sec:positive-domain-theorem}
\textit{Proposition.} For finite distinct $w_1,w_2\in(-1,1)$ and
$\alpha\ne0$, any solution of the sector-separated density system with
$\rho_1(1),\rho_2(1)>0$ has a unique maximal density-positive interval
$\mathcal I_\rho=(a_{\rho,\min},a_{\rho,\max})$, with
$0<a_{\rho,\min}<1<a_{\rho,\max}<\infty$. Its logarithmic width is
\begin{equation}
\ln\frac{a_{\rho,\max}}{a_{\rho,\min}}
=\frac{\left|\ln(\eta_+/\eta_-)\right|}{\gamma_+-\gamma_-}.
\label{eq:positive-domain-width}
\end{equation}
The width depends only on the equations of state, not on the initial
density ratio or $\alpha$. Provided $H^2(1)>0$, the admissible
nonstatic FLRW interval is the connected component containing $a=1$ of
$\mathcal I_\rho\cap\{a:H^2(a)>0\}$, and may be shorter.

\textit{Proof.} Set $\mathcal P(\gamma)=\gamma^2-C\gamma+2D$.
The coefficient definitions give
\begin{equation}
\begin{aligned}
\mathcal P\bigl(3(1+w_1)\bigr)&=3\beta(w_1-w_2),\\
\mathcal P\bigl(3(1+w_2)\bigr)&=3\zeta(w_1-w_2).
\end{aligned}
\label{eq:characteristic-at-dilutions}
\end{equation}
Since $A'(w)=4/(1+3w^2)^{3/2}>0$ and $\delta\pi_i>0$ in the open
square, both right-hand sides are negative. Therefore
\begin{equation}
\gamma_-<3(1+\min_i w_i)<3(1+\max_i w_i)<\gamma_+.
\label{eq:mode-straddling}
\end{equation}
For a nonzero modal vector $(d,f)$, conventional-sector conservation gives
$\gamma(d+f)=3[(1+w_1)d+(1+w_2)f]$. The case $d+f=0$ is impossible
for an unequal pair. If $d$ and $f$ had the same sign, or if one of them
vanished, $\gamma$ would lie in the closed interval between the two
conventional dilution exponents, contradicting
Eq.~\eqref{eq:mode-straddling}. Thus both eigenvectors have opposite-sign
components, and $\eta_\pm/\beta<0$; in particular, $\eta_+/\eta_->0$.

Positive initial densities require both modal coefficients $f_\pm$ to be
nonzero and of opposite signs. The individual density zeros satisfy
\begin{equation}
\begin{aligned}
a_{\rho_2=0}^{\Delta\gamma}&=-\frac{f_+}{f_-},\\
a_{\rho_1=0}^{\Delta\gamma}&=-\frac{\eta_+f_+}{\eta_-f_-},
\qquad \Delta\gamma=\gamma_+-\gamma_-.
\end{aligned}
\label{eq:individual-density-zero-scales}
\end{equation}
These are distinct, positive, finite scales. Outside the interval between
them one of the densities is negative; since both are positive at $a=1$,
that interval is $\mathcal I_\rho$. Dividing the two relations gives
Eq.~\eqref{eq:positive-domain-width}. This proves the statement for the
density system; a zero of $H^2$ can interrupt its cosmological realization.

For illustration, $a_{\rho,\max}/a_{\rho,\min}\simeq1.504$ for
$(w_1,w_2)=(0,1/3)$ and $\simeq2.755$ for $(0,-0.9)$. Changing the
initial density ratio shifts the endpoints in $\ln a$ without changing
the width. When one equation of state tends to $-1$ or $1$ with the other
fixed in the interior, the width diverges, consistently with the one-sided
endpoints of the vacuum and stiff families. The proposition does not include
these boundaries or the equal-equation-of-state diagonal.

A useful corollary is that no single density eigenmode has two positive
conventional densities in this open square. Consequently, on an admissible
nonstatic branch there, constant $q$ can occur only through rank-degenerate
gravitational cancellation, not through single-mode positive initial data.
Outside this domain the two mechanisms must still be distinguished, as in
Appendix~\ref{app:LD}. Section~\ref{sec:alphaparticular} gives the
corresponding gravitational eigenvector argument.

\subsection{Deceleration and the limits of background interpretation}
\label{sec:kinematics}
The general two-mode solution is not restricted to constant deceleration.
On a nonstatic interval with distinct real exponents,
Eq.~\eqref{eq:two-mode-H2} gives
\begin{equation}
q(a)=-1+
\frac{\gamma_-\chi_-a^{-\gamma_-}+\gamma_+\chi_+a^{-\gamma_+}}
{2\left(\chi_-a^{-\gamma_-}+\chi_+a^{-\gamma_+}\right)}.
\label{eq:q-two-mode}
\end{equation}
Differentiating with respect to $\ln a$ yields
\begin{equation}
\frac{\mathrm dq}{\mathrm d\ln a}
=-\frac{(\Delta\gamma)^2\chi_-\chi_+a^{-(\gamma_-+\gamma_+)}}
{2\left(\chi_-a^{-\gamma_-}+\chi_+a^{-\gamma_+}\right)^2}.
\label{eq:q-two-mode-derivative}
\end{equation}
Thus $q$ varies whenever both gravitational amplitudes are nonzero. For
$\chi_\pm>0$ it decreases as the universe expands. Under these
positive-amplitude conditions, if $\gamma_-<2<\gamma_+$, the formal change
from deceleration to acceleration occurs at
\begin{equation}
a_{\rm acc}=
\left[\frac{(\gamma_+-2)\chi_+}{(2-\gamma_-)\chi_-}
\right]^{1/\Delta\gamma}.
\label{eq:acceleration-scale}
\end{equation}
This is a kinematic condition, not an existence proof for an admissible
fluid history: $a_{\rm acc}$ must lie in the connected domain containing
$a=1$ on which both conventional densities and $H^2$ are positive. The
amplitudes are the combinations fixed by the original fluid and gravitational
equations, not independently adjustable fitting functions. Sign-indefinite
amplitudes or a zero of $H$ require the corresponding branch analysis.

The elementary vacuum-pair solution~\eqref{eq:hwv1} already illustrates the
nonconstant case. With $\chi_\pm>0$ and
$u\equiv(n_{\rm v}/2)\sqrt{\chi_-}(t-t_{\rm B})>0$, it gives
\begin{equation}
q(t)=-1+\frac{n_{\rm v}}{2}\operatorname{sech}^2u,
\label{eq:q-vacuum-hyperbolic}
\end{equation}
again restricted to the positive-density interval, rather than necessarily
extending to the formal $u\to0$ limit.

\paragraph{An admissible acceleration transition and its limits.}
\label{sec:admissible-transition}
The dust--vacuum pair provides an explicit transition without furnishing
a viable complete cosmology. For positive present-day densities and
$-1<\alpha<1/2$, $\alpha\ne0$, its amplitudes are positive and
$(\gamma_-,\gamma_+)=(0,9/2)$. Writing $1+z=a^{-1}$,
Eqs.~\eqref{eq:acceleration-scale} and
\eqref{eq:vacuum-branch-endpoint} give
\begin{equation}
\begin{aligned}
(1+z_{\rm acc})^{9/2}
 &=\frac{2(1-2\alpha)}{5(1+\alpha)}
 \left(1+\frac{3\rho_{\rm v0}}{\rho_{{\rm CDM}0}}\right),\\
(1+z_*)^{9/2}
 &=1+\frac{3\rho_{\rm v0}}{\rho_{{\rm CDM}0}}.
\end{aligned}
\label{eq:dust-vacuum-transition-endpoint}
\end{equation}
The transition is on the positive-density branch exactly when
$z_{\rm acc}<z_*$, or $-1/3<\alpha<1/2$ within this family, still excluding
$\alpha=0$. It may be in the future; requiring an already completed
transition adds $z_{\rm acc}>0$. For the illustrative ratio
$\rho_{\rm v0}/\rho_{{\rm CDM}0}=7/3$, the limit $\alpha\to0$ taken
\emph{within the nonzero-coupling family} gives $z_{\rm acc}\simeq0.295$
and $z_*\simeq0.587$. These are not the separately conserved
$\alpha=0$ GR values. The transition exists, but this positive branch
cannot be extrapolated to an early matter era; its matter dilution remains
$a^{-9/2}$ rather than $a^{-3}$.

For a general $w$--vacuum pair with $-1<w<1$, positive amplitudes, and
$K_w>0$, the deceleration at the vacuum-zero endpoint is
\begin{equation}
q_*\equiv\lim_{a\downarrow a_{{\rm v},*}}q(a)
=\frac{1+3w+\alpha[A(w)+3\sqrt{1+3w^2}]}
{2[1+\alpha A(w)]}.
\label{eq:vacuum-endpoint-deceleration}
\end{equation}
Since $q(a)$ decreases to $-1$, a deceleration-to-acceleration transition
lies strictly inside the branch iff $q_*>0$. The criterion $w>-1/3$
follows only in the $\alpha\to0$ closure limit; the finite-$\alpha$
condition is Eq.~\eqref{eq:vacuum-endpoint-deceleration}.
More generally, Eq.~\eqref{eq:char-discriminant-identity} implies
$\gamma_+-\gamma_->3$: the unequal-EoS, nonzero-$\alpha$, two-fluid
closure cannot reproduce exactly the two nonzero $\Lambda$CDM terms with
exponents $(0,3)$. This is not a bound on approximate fits over a finite
redshift interval, nor a statement about models with additional species.

By contrast, retaining one gravitational mode gives a constant $q$. For a
power-law expanding interval,
\begin{equation}
a(t)\propto(t-t_{\rm B})^p,\qquad
H=\frac{p}{t-t_{\rm B}},\qquad q=\frac1p-1,
\label{eq:power-law-kinematics}
\end{equation}
with $p>0$ and $t>t_{\rm B}$. A single such law cannot describe both a
preceding decelerating epoch and the later accelerating epoch, whose
transition is supported by supernova observations~\cite{SupernovaSearchTeam:2004lze}.
An inverse-time Hubble rate is not problematic merely as an individual
expansion regime; the limitation is its extrapolation with one fixed exponent
across the full cosmological history. Indeed, power-law fits to some
low-redshift expansion data need not supply a consistent early-to-late history,
as the comparison with primordial-nucleosynthesis requirements in
Ref.~\cite{Kumar:2011sw} illustrates.

The rank-degenerate solutions of Sect.~\ref{sec:alphaparticular} expose this
limitation directly. Equation~\eqref{eq:vacuum-power-q} has $q\geq5/4$ and
never accelerates; Eq.~\eqref{eq:stiff-like-H} gives $q=2$; and
Eq.~\eqref{eq:accelerated-power-q} has $-1<q\leq-1/4$ for $-1<w<1$, with
no preceding decelerating phase on the same branch. The de Sitter solution
has $q=-1$ throughout.
Accordingly, when extrapolated as global constant-parameter cosmologies, the
fixed-exponent inverse-time histories in Eqs.~\eqref{eq:vacuum-power-H},
\eqref{eq:stiff-like-H}, and \eqref{eq:accelerated-power-H} are incompatible
with the observed expansion history: their constant deceleration parameters
cannot reproduce the transition from an earlier decelerating epoch to
late-time acceleration~\cite{SupernovaSearchTeam:2004lze}.
These backgrounds are retained to display exact modal
cancellation and its physical-domain restrictions, not as complete
alternatives to $\Lambda$CDM. Patching different solutions together would
require new matching conditions and dynamics, neither of which is supplied
by the present constant-parameter closure.

Even a two-mode history with an admissible acceleration transition is not
thereby an observationally viable cosmology. The dust--vacuum example has
$\rho_{\rm CDM}\propto a^{-9/2}$ for every $\alpha\ne0$,
and the proposition in Sect.~\ref{sec:nvp} proves finite density-positive
width for distinct $w_i\in(-1,1)$. The boundary vacuum and stiff families
have the one-sided endpoints discussed in Sects.~\ref{sec:csga1} and
\ref{sec:wstiff}.%
 Small $|\alpha|$ does not control these effects. The
present work supplies neither a likelihood analysis nor a covariant
perturbation model. A realistic interpretation would require suitable
background epochs and domains, a covariant specification of the currents and
matter response, a stability analysis, and confrontation with expansion and
structure-growth observations.%

\section{\texorpdfstring{$\alpha$}{alpha}-dependent rank degeneracies and modal cancellations}
\label{sec:alphaparticular}

Recall $K_i\equiv1+\alpha A_i$. Since each density obeys
Eqs.~\eqref{eq:dr1}--\eqref{eq:dr2}, their constant linear combination
\begin{equation}
y(a)\equiv H^2(a)=\frac{\kappa}{3}
\left[K_1\rho_1(a)+K_2\rho_2(a)\right]
\end{equation}
obeys, for $\alpha\ne0$ within the sector-separated closure family,
\begin{equation}
y''+C(w_1,w_2)y'+2D(w_1,w_2)y=0,
\label{eq:linear-H2}
\end{equation}
where a prime denotes differentiation with respect to $\ln a$. On every
interval with $H\neq0$, $y'=2\dot H$ and $y''=2\ddot H/H$. Hence
\begin{equation}
\ddot H+C(w_1,w_2)H\dot H+D(w_1,w_2)H^3=0.
\label{eq:mstro}
\end{equation}
This modified Li\'enard equation follows from the density evolution throughout
that nonzero-$\alpha$ family, including the rank-degenerate values considered below; it is
a member of the broader families studied in
Refs.~\cite{Chimento:1997uj,2014JEnMa..89..193H}.

\paragraph{Exact integration of the Li\'enard equation.}
The preceding transformation also solves Eq.~\eqref{eq:mstro}. Set
$x=\ln a$ on a connected interval with $H\ne0$ and $y(x)=H^2$. The identities
$y'=2\dot H$ and $y''=2\ddot H/H$ convert it exactly into the
constant-coefficient linear equation~\eqref{eq:linear-H2}. For distinct real
roots its solution is
\begin{equation}
y(x)=\chi_-e^{-\gamma_-x}+\chi_+e^{-\gamma_+x},\qquad
\gamma_\pm=\frac{C\pm\sqrt{C^2-8D}}{2}.
\label{eq:Lienard-exact-y}
\end{equation}
Together with $H(a)=\pm\sqrt{y(\ln a)}$ and
$\mathrm dt=\mathrm da/(aH)$, this is an exact parametric integration.
Equations~\eqref{eq:ggquadrature}--\eqref{eq:gghs} give its expanding
cosmic-time quadrature and, where the stated conditions hold, its
hypergeometric representation. An elementary inversion to $a(t)$ is not
required for exact solvability. This integrable structure and its
interacting-fluid antecedent are not new~\cite{Barrow:2006hia,Chimento:1997uj};
here the coefficients and admissible amplitudes are fixed by the declared
closure and gravitational equations.

Elementary cosmic-time solutions follow immediately. For a nonzero real
root $\gamma$ of $\gamma^2-C\gamma+2D=0$, a single mode gives
\begin{equation}
\begin{aligned}
H(t)&=\frac{H_0}{1+\frac{\gamma}{2}H_0(t-t_0)},\\
a(t)&=\left[1+\frac{\gamma}{2}H_0(t-t_0)\right]^{2/\gamma},
\end{aligned}
\label{eq:Lienard-power}
\end{equation}
with a positive bracket and $H_0>0$ on the expanding interval. Equivalently,
substitution of $H=k/(t-t_{\rm B})$ gives $Dk^2-Ck+2=0$, with
$k=2/\gamma$. A nonzero constant $H$ solves Eq.~\eqref{eq:mstro} precisely
when $D=0$.

When $D=0$, there is also the first integral
\begin{equation}
\dot H+\frac C2H^2=\mathcal I,
\label{eq:Lienard-first-integral}
\end{equation}
where $\mathcal I$ is constant. The unequal-EoS zero set in
Eq.~\eqref{eq:D-zero-locus} has $C>0$. For $\mathcal I=Ch^2/2>0$, one
expanding solution is
\begin{equation}
\begin{aligned}
H(t)&=h\coth u,\\
a(t)&=\left[\frac{\sinh u}{\sinh u_0}\right]^{2/C},\qquad
u=\frac{Ch}{2}(t-t_{\rm B})>0,
\end{aligned}
\label{eq:Lienard-coth}
\end{equation}
where $u_0=u(t_0)$ and $h>0$. The vacuum-pair solution
\eqref{eq:wv-hyperbolic-a}--\eqref{eq:hwv1} is recovered with
$C=n_{\rm v}$ and $h=\sqrt{\chi_-}$. The same first integral admits the
$t>t_{\rm B}$ solution $H=h\tanh u$,
$a=[\cosh u/\cosh u_0]^{2/C}$ and the equilibrium $H=h$; the
$H=-h$ equilibrium is contracting. These are mathematical solutions of the
reduced equation. Their fluid realization still requires the original
Friedmann, pressure, and conservation equations and positive conventional
densities.
The reduced equation alone allows both $\chi_\pm$ to be nonzero;
at a rank-degenerate coupling the full system forces one gravitational
amplitude to vanish.%

\paragraph{Rank loss and the status of the special backgrounds.}
The constant-$q$ power laws below arise when the gravitational combination
loses one mode; they are not the general solution of the Li\'enard equation.
They will be used as exact consistency diagnostics, subject to the
cosmological limitations established in Sect.~\ref{sec:kinematics}.%

A separate issue arises when the two densities are reconstructed from the
Friedmann and pressure equations. The coefficient matrix $\mathbf X$ defined in
Appendix~\ref{app:LD} has determinant
\begin{equation}
\det\mathbf X=3H\left[(w_1-w_2)+E(w_1,w_2)\right],
\label{eq:detx-main}
\end{equation}
where
\begin{align}
\label{eq:Edef}
&E(w_1,w_2)=\alpha^2\left[(1+w_1)A_2B_1-(1+w_2)A_1B_2\right]\nonumber \\
&\quad+\alpha\left[(1+w_1)(A_2+B_1)-(1+w_2)(A_1+B_2)\right].
\end{align}
For a nonstatic background, the condition
\begin{align}
\label{eq:extra}
(w_1-w_2)+E(w_1,w_2)=0
\end{align}
does not define a second dynamical branch of Eq.~\eqref{eq:mstro}; it makes the
density-reconstruction matrix rank deficient. Its solutions must be substituted
into the original Friedmann, pressure, and conservation equations. They select
modal-cancellation subfamilies of the same Li\'enard solution family.
\paragraph{Physical interpretation of rank loss.}
\label{sec:rank-interpretation}
Let $R_i\equiv K_i+P_i=(1+w_i)(1+\alpha B_i)$, so that adding the
Friedmann and pressure equations gives
$-2\dot H=\kappa(R_1\rho_1+R_2\rho_2)$. On a nonstatic interval,
$\det\mathbf X=3H(R_1K_2-R_2K_1)$. For a finite unequal pair,
$\boldsymbol K=(K_1,K_2)^T$ cannot vanish identically, since
$A_1\ne A_2$. Rank loss is therefore equivalent to the density-ratio-independent condition
\begin{equation}
P_i=\widetilde w\,K_i\quad(i=1,2),\qquad
R_i=(1+\widetilde w)K_i
\label{eq:rank-common-eos}
\end{equation}
for one finite constant $\widetilde w$. If both $K_i\ne0$, the two
EMSF-dressed species have the same pressure-to-density ratio $P_i/K_i$.
If one $K_i=0$, its $P_i$ also vanishes: that dressed species is
gravitationally null and has no defined individual ratio. In either case,
on any admissible nonstatic background,
\begin{equation}
\frac{\dot H}{H^2}=-\frac32(1+\widetilde w),\qquad
q=\frac12(1+3\widetilde w).
\label{eq:rank-constant-q}
\end{equation}
The cancellation is independent of the density initial conditions, but
those conditions must still admit $H^2>0$ and positive conventional densities.

To see why precisely one gravitational mode survives, write
$\boldsymbol\rho'=-\mathbf M\boldsymbol\rho$, where
\begin{equation}
\mathbf M=\begin{pmatrix}
3(1+w_1)-\beta & -\zeta\\
\beta & 3(1+w_2)+\zeta
\end{pmatrix}.
\label{eq:density-evolution-matrix}
\end{equation}
The total conservation equation gives
\begin{equation}
\boldsymbol K^T\mathbf M=3\boldsymbol R^T.
\label{eq:gravitational-left-eigenvector}
\end{equation}
At rank loss, $\boldsymbol K^T$ is consequently a left eigenvector with
eigenvalue $3(1+\widetilde w)\in\{\gamma_-,\gamma_+\}$. Its contraction
with the other, distinct right eigenvector vanishes, removing that mode
from $H^2=\kappa\boldsymbol K^T\boldsymbol\rho/3$. Conversely, for a
solution in which both density modes are excited, losing one gravitational
mode makes $\boldsymbol K^T$ a left eigenvector and enforces
Eq.~\eqref{eq:rank-common-eos}. Combined with the positive-density
proposition in Sect.~\ref{sec:nvp}, this proves the
constant-$q$ corollary for distinct $w_i\in(-1,1)$.
The four examples below have $\widetilde w=w_{\rm v,eff},-1,1$, and
$w_{\rm s,eff}$, respectively.

Substitution of Eq.~\eqref{eq:Edef} into Eq.~\eqref{eq:extra} gives
\begin{equation}
\alpha^2s(w_1,w_2)+\alpha p(w_1,w_2)+(w_1-w_2)=0, 
\label{alphaquad}
\end{equation}
where the functions $s(w_1,w_2)$ and $p(w_1,w_2)$ are defined as:
\begin{align}
   \label{eq:s}
    s(w_1,w_2)=&\frac{4(w_1-w_2)(3w_1w_2-1)}{\sqrt{(3w_1^2+1)(3w_2^2+1)}},\\
    p(w_1,w_2)=&\frac{3w_1^2-4w_1w_2+1}{\sqrt{1+3w_1^2}}-\frac{3w_2^2-4w_1w_2+1}{\sqrt{1+3w_2^2}}.
    \label{eq:p}
\end{align}
Equation~\eqref{alphaquad} is genuinely quadratic only when $s(w_1,w_2)\neq0$. Defining
\begin{equation}
\label{eq:Delta}
\Delta(w_1,w_2)\equiv
p(w_1,w_2)^2-4s(w_1,w_2)(w_1-w_2),
\end{equation}
the two real algebraic roots for $s\neq0$ and $\Delta\geq0$ are
\begin{align}
\alpha_{\pm}=&-\frac{p(w_1,w_2)\pm\sqrt{\Delta(w_1,w_2)}}{2s(w_1,w_2)}.
\label{eq:aalp}
\end{align}
The labels $+$ and $-$ distinguish the two algebraic roots and do not, in
general, indicate the sign of $\alpha$.

\paragraph{Global sign of the discriminant.}
The reality of the roots can be decided analytically. Define
\begin{equation}
u_i\equiv\arctan(\sqrt3\,w_i),\qquad
m\equiv\frac{u_1+u_2}{2},\qquad
d\equiv\frac{u_1-u_2}{2},
\end{equation}
and set $M\equiv\sin^2m$ and
$\mathcal E\equiv\cos u_1\cos u_2$. Direct substitution into
Eq.~\eqref{eq:Delta} gives the exact factorization
\begin{align}
\Delta={}&\frac{4\sin^2d}
{9\cos^2u_1\cos^2u_2}\Big[M(8M-7)^2\nonumber\\
&\quad+8\mathcal E(8M^2-13M+6)+16\mathcal E^2M\Big].
\label{eq:Delta-factorization}
\end{align}
For finite real $w_i$, one has $\mathcal E>0$ and $0\leq M<1$.
Moreover, $8M^2-13M+6$ is strictly positive because its discriminant is
$-23$. Every term in the bracket is therefore nonnegative and the middle term
is strictly positive. Since $\arctan$ is injective,
\begin{equation}
\Delta(w_1,w_2)>0\quad\Longleftrightarrow\quad w_1\ne w_2,
\qquad
\Delta(w,w)=0.
\label{eq:Delta-theorem}
\end{equation}
Consequently, every genuinely quadratic case $s\ne0$ has two distinct real
rank-degenerate couplings; no finite unequal-EoS region with complex roots
exists.

\begin{figure}[!htbp]
\centering
\includegraphics[width=\columnwidth]{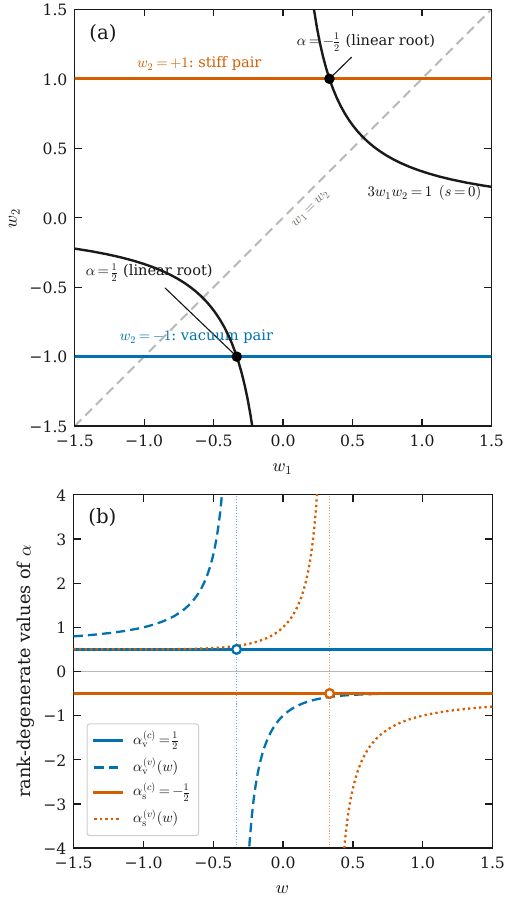}
\caption{Rank-degenerate subfamilies. \textbf{a}~The curves $3w_1w_2=1$ are the
linear locus $s=0$; the equal-EoS diagonal is excluded. The horizontal lines
are the vacuum and stiff families, and the dots mark the points where the
finite linear roots are $\alpha=\pm\tfrac12$. The equal-EoS intersections $(-1,-1)$ and
$(1,1)$ are excluded. \textbf{b}~Each family has a constant root
$\alpha=\pm\tfrac12$ and a variable root, which diverges at
$w=\mp\tfrac13$ (dotted lines); the open circles show the surviving finite
linear roots. The plotted curves are formal continuations through the isolated
equal-EoS points $w=-1$ and $w=1$. At those points
Eq.~\eqref{alphaquad} is an identity for every $\alpha$, not a two-root
condition.}
\label{fig:degalpha}
\end{figure}

When $s=0$ but $w_1\ne w_2$, Eq.~\eqref{eq:Delta-theorem} implies
$p\ne0$. Equation~\eqref{alphaquad} is then linear and admits the single finite solution
\begin{equation}
\alpha=-\frac{w_1-w_2}{p(w_1,w_2)}.
\label{eq:a0}
\end{equation}
From Eq.~\eqref{eq:s}, $s=0$ occurs on the equal-EoS diagonal and on
$3w_1w_2=1$. The unequal-EoS locus $3w_1w_2=1$, on which
Eq.~\eqref{alphaquad} becomes linear in $\alpha$, and its intersections with
the vacuum and stiff families are shown in Fig.~\ref{fig:degalpha}a. On the diagonal one also has $p=0$, so
Eq.~\eqref{alphaquad} is an identity and the original equations must be used.
At unequal points of $3w_1w_2=1$, Eq.~\eqref{eq:a0} gives the unique linear
root; Eq.~\eqref{eq:aalp} must not be used. For example, the pairs
$(w_1,w_2)=(-\tfrac13,-1)$ and $(\tfrac13,1)$ lead respectively to
$\alpha=\tfrac12$ and $\alpha=-\tfrac12$.

For clarity, we label the roots in the two one-parameter families by their
functional character rather than by the ordering-dependent signs in
Eq.~\eqref{eq:aalp}:
\begin{align}
\alpha_{\rm v}^{(c)}&\equiv\frac12,&
\alpha_{\rm v}^{(v)}(w)&\equiv-\frac{\sqrt{1+3w^2}}{1+3w},
\qquad (w_2=-1),\nonumber\\
\alpha_{\rm s}^{(c)}&\equiv-\frac12,&
\alpha_{\rm s}^{(v)}(w)&\equiv+\frac{\sqrt{1+3w^2}}{1-3w},
\qquad (w_2=1).
\label{eq:family-root-names}
\end{align}
Here $(c)$ denotes the species-independent root and $(v)$ the variable root.
Both families are displayed in Fig.~\ref{fig:degalpha}b. At
$w=\mp\tfrac13$ the variable root diverges and only the constant linear root
survives. At $w=-1$ in the vacuum family and $w=1$ in the stiff family, the
two species have equal EoS; Eq.~\eqref{alphaquad} is then an identity for every
$\alpha$, so the displayed two-root formulae do not apply.

The linear-root and equal-EoS cases must be checked in the original background
equations. When Eq.~\eqref{eq:extra} holds, the density-reconstruction matrix is
singular, but the Li\'enard equation~\eqref{eq:mstro} still follows from the
density evolution. Equation~\eqref{eq:extra} therefore selects a rank-deficient
parameter subfamily rather than a separate dynamical branch.

Equations~\eqref{eq:edpt3} and \eqref{eq:edpt4} show that $\alpha$ fixes the
algebraic EMSF amplitude relative to each conventional source. The general
quadratic roots~\eqref{eq:aalp} can equivalently be written as present-day
density-parameter ratios:
\begin{equation}
\begin{aligned}
\left(\frac{\Omega^{\rm EMSF}_{10(20)}}
{\Omega_{10(20)}}\right)_{\pm}
={}&-\frac{p\pm\sqrt{p^2-4s(w_1-w_2)}}{2s}\\
&\times\frac{4w_{1(2)}}{\sqrt{1+3w^2_{1(2)}}}.
\end{aligned}
\label{eq:Oaalp}
\end{equation}
Thus, this relation applies only to the genuinely quadratic case ($s\neq0$), while Eq.~\eqref{eq:a0} must be used when the degeneracy condition is linear.
This relationship between $\Omega^{\rm EMSF}$ and $\Omega$ reduces the number
of free background parameters. Appendix~\ref{app:LD} distinguishes this
coefficient-matrix degeneracy from linear dependence of the two density
solutions. Otherwise, the EMSF amplitude relative to the conventional source
remains free, as in Sect.~\ref{sec:cossol}.

We now restrict the general-$\alpha$ solutions of Sect.~\ref{sec:cossol} to the
roots of Eq.~\eqref{eq:extra} and substitute their parameter and density
relations into the original Friedmann equations. The resulting solutions are
rank-deficient subfamilies of the original background system.

\subsection{Power-law universes with a perfect fluid and vacuum energy}

We consider a conventional source with $w_1=w={\rm const.}\ne-1$ and vacuum
energy ($w_2=-1$) at the species-independent root
$\alpha=\alpha_{\rm v}^{(c)}=1/2$. At $w=-1/3$ this is the unique
finite linear root of Eq.~\eqref{eq:a0}. The corresponding present-day density
relations are
\begin{equation}
\begin{aligned}
\left(\frac{\Omega_{w0}^{\rm EMSF}}{\Omega_{w0}}\right)_{c}
&=\frac{w}{w_{\rm v,eff}(w)},\\
\left(\frac{\Omega_{{\rm v}0}^{\rm EMSF}}{\Omega_{{\rm v}0}}\right)_{c}
&=-1,
\end{aligned}
\label{eq:omrel1}
\end{equation}
where the second relation implies that the vacuum energy and its EMSF partner
cancel in the Friedmann equation.

Substitution of Eq.~\eqref{eq:omrel1} into the Friedmann
equation~\eqref{eq:wvfr1} gives
\begin{equation}
\begin{aligned}
\frac{H^2}{H_0^2}={}&\Omega_{w0}
\frac{w_{\rm v,eff}(w)+w}{w_{\rm v,eff}(w)}\\
&\times a^{-3[1+w_{\rm v,eff}(w)]}.
\end{aligned}
\label{eq:wa1}
\end{equation}
The scale factor is
\begin{equation}
a(t)=\left[\frac{n_{\rm v}}{2}H_0
\sqrt{\frac{w_{\rm v,eff}(w)+w}{w_{\rm v,eff}(w)}
\Omega_{w0}}\,t\right]^{2/n_{\rm v}}.
\label{eq:cvh}
\end{equation}
The corresponding Hubble parameter is
\begin{equation}
\begin{aligned}
H(t)=\frac{2}{n_{\rm v}t},
\label{eq:vacuum-power-H}
 \end{aligned}
\end{equation}
and the deceleration parameter $q=-1-\dot H/H^2$ can be written as
\begin{equation}
\begin{aligned}
    q=\frac{1}{2}+\frac{3}{4}\sqrt{3w^2+1},
\label{eq:vacuum-power-q}
 \end{aligned}
\end{equation}
which is positive for all real $w$. For $\Omega_{w0}>0$, the expanding
solution requires $w>-1$ and is decelerating. At this root the vacuum and its
EMSF partner cancel in the Friedmann equation, leaving the remaining mode to
control the expansion. For $-1<w<1$, however, the solution is valid only for
$a>a_{{\rm v},*}$ as defined in
Eq.~\eqref{eq:vacuum-branch-endpoint}; its formal $a\to0$ limit lies outside
the positive-density branch.

The origin of time in Eq.~\eqref{eq:vacuum-power-H} is chosen so that
$t_{\rm B}=0$; a shift restores $H=2/[n_{\rm v}(t-t_{\rm B})]$. This
inverse-time law and its strictly positive, constant $q$ cannot describe the
late accelerating epoch. The result is a cancellation example on its stated
domain, not a fit to the observed expansion history.%

\subsection{Steady-state-like de Sitter universe with a perfect fluid and vacuum energy}
\label{sec:de-sitter}
The second vacuum root is
$\alpha=\alpha_{\rm v}^{(v)}(w)=-\sqrt{1+3w^2}/(1+3w)$, with $w\ne-1$.
It has no finite continuation at $w=-1/3$, where only
$\alpha_{\rm v}^{(c)}=1/2$ survives. The corresponding density relations are
\begin{equation}
\begin{aligned}
\left(\frac{\Omega_{w0}^{\rm EMSF}}{\Omega_{w0}}\right)_{v}
&=-\frac{4w}{1+3w},\\
\left(\frac{\Omega_{{\rm v}0}^{\rm EMSF}}{\Omega_{{\rm v}0}}\right)_{v}
&=\frac{4w_{\rm v,eff}(w)}{1+3w}.
\end{aligned}
\label{eq:omrel2}
\end{equation}
Substitution of Eq.~\eqref{eq:omrel2} into the Friedmann
equation~\eqref{eq:wvfr1} gives the constant Hubble parameter
\begin{equation}
\begin{aligned}
H={}&\pm H_0\left(1+\frac{4w_{\rm v,eff}}{1+3w}\right)^{1/2}\\
&\times\left(\Omega_{\rm v0}
-\frac{w-w_{\rm v,eff}}{1+w_{\rm v,eff}}\Omega_{w0}\right)^{1/2},
\end{aligned}
    \label{eq:Hconst}
\end{equation}
where the square-root expression equals unity by Eq.~\eqref{eq:fv}. Hence
\begin{equation}
H=\pm H_0, \qquad a(t)=e^{H(t-t_0)}.
\label{eq:wvs22}
\end{equation}
The effective gravitational density is constant even though the conventional
sector contains a $w\neq-1$ component. For nonnegative conventional density
parameters with $-1\leq w\leq1$, the positive-$H^2$ realization requires
$w>-1/3$. This is a de Sitter, steady-state-like background related to the EMSG
construction of Ref.~\cite{Akarsu:2023nyl}. It is not past eternal within the
positive-density matter branch: for $-1/3<w<1$, the vacuum density reaches
zero at $a_{{\rm v},*}$, so Eq.~\eqref{eq:wvs22} describes only the segment
$a>a_{{\rm v},*}$ unless a separate continuation is constructed.

\subsection{Stiff-fluid-like expansion with a perfect fluid}
\label{sec:stiffdom}
For a conventional source with $w_1=w={\rm const.}\ne1$ and a stiff fluid
($w_2=1$), consider the root
$\alpha=\alpha_{\rm s}^{(v)}(w)=\sqrt{1+3w^2}/(1-3w)$.
This root has no finite continuation at $w=1/3$, where the only finite solution
is $\alpha_{\rm s}^{(c)}=-1/2$. The density relations are
\begin{equation}
\label{eq:emsfusual}
\begin{aligned}
\left(\frac{\Omega_{w0}^{\rm EMSF}}{\Omega_{w0}}\right)_{v}
&=\frac{4w}{1-3w},\\
\left(\frac{\Omega_{{\rm s}0}^{\rm EMSF}}{\Omega_{{\rm s}0}}\right)_{v}
&=-\frac{4w_{\rm s,eff}(w)}{1-3w}.
\end{aligned}
\end{equation}
Substituting these relations into the Friedmann equation \eqref{eq:gwsfh}, we obtain
\begin{equation}
\begin{aligned}
\frac{H^2}{H_0^2}={}&\left[\frac{1+w}{1-3w}\Omega_{w0}
+\left(1-\frac{4w_{\rm s,eff}}{1-3w}\right)\Omega_{\rm s0}\right]\\
&\times a^{-6},
\end{aligned}
\label{eq:friedstiff}
\end{equation}
and Eq.~\eqref{eq:fvstiff} sets the square bracket to unity, giving
\begin{equation}
\begin{aligned}
H^2=H_0^2a^{-6},\quad \textnormal{and}\quad a(t)=\left[1+3H_0(t-t_0)\right]^{1/3}.
\end{aligned}
\end{equation}
The corresponding Hubble parameter is
 \begin{equation}
\begin{aligned}
H(t)=\frac{H_0}{1+3H_0(t-t_0)}.
\label{eq:stiff-like-H}
\end{aligned}
\end{equation}
Within this branch, the combined gravitational contribution scales exactly as a
stiff component. For nonnegative conventional density parameters and
$-1\leq w\leq1$, the positive-$H^2$ realization exists for $w<1/3$; the
value $w=1/3$ is excluded. For $-1<w<1/3$, the conventional stiff density
nevertheless reaches zero at $a_{{\rm s},*}$, so this solution does not define
a global $a\to\infty$ cosmology on the assumed positive-density branch.

In particular, Eq.~\eqref{eq:stiff-like-H} is
$H=1/[3(t-t_{\rm B})]$ with $t_{\rm B}=t_0-1/(3H_0)$ and $q=2$.
It neither accelerates nor supplies the usual matter-dominated expansion;
it cannot serve as a complete observed cosmological history. Its purpose is
to identify which gravitational mode survives the cancellation.%

\subsection{Accelerated power-law subfamily with a stiff fluid}

For the same unequal-EoS sources, the species-independent root
$\alpha=\alpha_{\rm s}^{(c)}=-1/2$ gives a power-law subfamily. At
$w=1/3$ it is the unique finite linear root, and the density relations remain
regular when evaluated in the original equations. They are
\begin{equation}
\begin{aligned}
\left(\frac{\Omega_{w0}^{\rm EMSF}}{\Omega_{w0}}\right)_{c}
&=\frac{w}{w_{\rm s,eff}(w)},\\
\left(\frac{\Omega_{{\rm s}0}^{\rm EMSF}}{\Omega_{{\rm s}0}}\right)_{c}
&=-1,
\end{aligned}
\label{eq:ostp}
\end{equation}
where the second relation implies that the stiff fluid and its EMSF partner
cancel in the Friedmann equation. Substitution of Eq.~\eqref{eq:ostp} into
Eq.~\eqref{eq:gwsfh} gives
\begin{equation}
\begin{aligned}
\frac{H^2}{H_0^2}={}&\Omega_{w0}
\frac{w_{\rm s,eff}(w)+w}{w_{\rm s,eff}(w)}\\
&\times a^{-3[1+w_{\rm s,eff}(w)]}.
\end{aligned}
\end{equation}
For $-1<w<1$, the normalized expanding solution is
\begin{equation}
a(t)=\left[1+\frac{3}{2}\left(1+w_{\rm s,eff}\right)
H_0(t-t_0)\right]^{\frac{2}{3(1+w_{\rm s,eff})}},
\label{eq:stiff-power-normalized}
\end{equation}
with
\begin{equation}
H(t)=\frac{H_0}{1+\frac{3}{2}(1+w_{\rm s,eff})H_0(t-t_0)}.
\label{eq:accelerated-power-H}
\end{equation}
The deceleration parameter is
\begin{equation}
\begin{aligned}
    q=\frac{1}{2}-\frac{3}{4}\sqrt{3w^2+1},
\label{eq:accelerated-power-q}
 \end{aligned}
\end{equation}
so $-1<q\leq-1/4$ for $-1<w<1$. Dust and radiation examples are
therefore accelerated even though either source would decelerate in the
corresponding conventional model. The solution is only local: for
$-1<w<1$, the stiff density reaches zero at $a_{{\rm s},*}$, and the formal
$a\to\infty$ power-law limit lies outside the positive-density branch. The
endpoint $w=-1$ must be treated separately; it has
$w_{\rm s,eff}=-1$ and gives the de Sitter solution
$a(t)=\exp[H_0(t-t_0)]$. For $w<-1$, the formal superaccelerated solution is
instead written with $t_{\rm s}-t$ and reaches a big-rip singularity at finite
$t_{\rm s}$~\cite{Caldwell:2003vq}, subject again to the density-domain
constraints.

Conversely, the persistent acceleration in
Eqs.~\eqref{eq:accelerated-power-H}--\eqref{eq:accelerated-power-q} does
not include a transition from an earlier decelerating epoch. Its inverse-time
form after shifting the time origin, together with its finite positive-density
domain, precludes interpreting this single subfamily as a complete
$\Lambda$CDM alternative. Constant negative $q$ is an exact feature of the
surviving mode, not evidence of observational viability.%

\section{Conclusions}
\label{sec:conc}

This analysis separates three inputs that are often conflated in multi-fluid
matter-type gravity: conservation of the total effective stress tensor,
constituent currents selected by the matter equations of a complete action, and
additional conservation assignments imposed as phenomenological closures. The
twice-contracted Bianchi identity determines only the first. Moreover, the
division of the total matter stress tensor into ``conventional'' and
``modification'' pieces depends on the chosen decomposition of the total matter
Lagrangian; only the total tensor is invariant under a reshuffling of that
split. For a minimal species-separable action, the independently
diffeomorphism-invariant matter equations conserve each conventional species
together with its own modification partner. The sector-separation rule studied
here instead conserves the sum of the conventional tensors and the sum of the
modification tensors separately. We have accordingly distinguished covariant
sub-sector currents from the scalar kernels obtained only after homogeneous
projection and algebraic diagonalization, and have displayed representative
assignments at both levels.

The case study combines this sector-separation rule with a common coupling
$\alpha_1=\alpha_2\equiv\alpha$ and an algebraic perfect-fluid prescription in
which the contracted metric-Hessian contribution in
Eq.~\eqref{eq:fluid-hessian-prescription} is set to zero. Neither the
metric-variation prescription nor the sector-separated current assignment is
derived here from an explicit off-shell fluid action. The resulting
construction is therefore an EMSG-structured background closure rather than
the on-shell cosmology selected by the minimal separable EMSG action. This
qualification does not invalidate the background mathematics, but it fixes the
level at which the results can be interpreted.

The modification is species separable, excluding the mixed contractions of
a generic function of the total conventional $T_{\mu\nu}T^{\mu\nu}$. Both
the rank-degeneracy map and the interaction kernel are conditional on this restriction. In particular, setting the contracted Hessian to zero is a
diagnostic algebraic choice, not a demonstrated consequence of an off-shell
fluid action.%

The explicit constant-$w$ Brown and regular Schutz comparisons give
$T^{\rm mod}_{\mu\nu}=\alpha k_iT_{\mu\nu}$ and action-selected separate
conservation whenever $1+\alpha k_i\ne0$ (see Table~\ref{tab:attribution}).
Sector separation alone also
implies zero exchange when $k_1\ne k_2$, but is underdetermined when
$k_1=k_2$. These comparisons neither supply a completion of the algebraic
case study nor exclude other covariant actions. In that case study,
Eqs.~\eqref{eq:pressure-mismatch}--\eqref{eq:kernel-mismatch-current}
identify the interaction as compensation for the partner's pressure
mismatch; cancellation of the common coupling explains its
$\alpha$ independence.
For constant-$w$ fluids, the action comparisons also show that modified
dilution laws obtained with the algebraic stress are tied to that
prescription.%

For finite unequal EoS parameters and every nonzero common coupling,
diagonalization gives a Barrow--Clifton-type two-fluid system~\cite{Barrow:2006hia}. Its transfer
coefficients are fixed by $(w_1,w_2)$ rather than introduced as independent
phenomenological parameters. This fixing is nevertheless a property of the
complete algebraic closure, not of the EMSG action alone. In particular, the
modification-sector conservation condition has the schematic form
$\alpha\mathcal F=0$. Dividing by $\alpha$ makes the resulting transfer
coefficients independent of $\alpha$, whereas at $\alpha=0$ that condition
loses its content. At exactly $\alpha=0$, the closure supplies only conservation
of the total conventional tensor; separate conservation of uncoupled GR species
follows after their matter equations are reinstated. 
The generic nonvanishing-exchange limit taken inside the nonzero-coupling
family therefore differs from uncoupled GR. This discontinuity belongs to
the sector-separated conservation closure acting on the algebraic stress
(Table~\ref{tab:attribution}), not to EMSG itself. The
stiff--vacuum pair, for which both transfer coefficients vanish, retains the
uncoupled dilution laws and approaches its GR background continuously, as
Eq.~\eqref{eq:stiff-vacuum-GR-limit} makes explicit.%

The conventional densities generally share the two characteristic modes
$a^{-\gamma_\pm}$, and their gravitational combination obeys the modified
Li\'enard equation~\eqref{eq:mstro} on intervals where $H\ne0$. Both exponents
are nonzero only when $D\ne0$. Besides the vacuum-energy cases, $D=0$ on the
unequal nonvacuum locus
$1+w_1+w_2-3w_1w_2=0$ with $w_1,w_2>1/3$; there the modes are a constant and
$a^{-C}$, with $C>6$. Whenever the constant mode is present, its two density
amplitudes have opposite signs. Rank loss in the map from the two densities to the Friedmann and
pressure equations is a separate phenomenon. It selects modal-cancellation
subfamilies and does not generate a second dynamical branch.

The transformation to $H^2(\ln a)$ also provides the exact parametric
integration of the Li\'enard equation. Its elementary cosmic-time solutions
include the power laws~\eqref{eq:Lienard-power}, constant-$H$ solutions for
$D=0$, and the hyperbolic family~\eqref{eq:Lienard-coth}, with admissibility
always checked in the original equations.%

For the common-coupling degeneracy polynomial, the exact factorization
\eqref{eq:Delta-factorization} proves that the discriminant is strictly positive
for every finite $w_1\ne w_2$. Every genuinely quadratic case therefore has two
distinct real rank-degenerate couplings. At unequal points of
$3w_1w_2=1$, the quadratic coefficient vanishes and one finite linear root
remains. On the equal-EoS diagonal the degeneracy equation is instead an
identity, and the original background equations must be used.

The density-ratio-independent condition $P_i=\widetilde wK_i$ identifies the physical
meaning of rank loss, including the possibility of a gravitationally null
dressed species. The gravitational coefficient vector is then a left
eigenvector of the density evolution matrix, so precisely one mode
survives on a nonstatic admissible background. This is why the special
solutions have constant $q$, independently of their density initial ratio;
positivity and $H^2>0$ remain separate requirements.%

The vacuum and stiff examples make the cancellation mechanism explicit. At
$\alpha_{\rm v}^{(c)}=1/2$, vacuum energy cancels its EMSF partner in the
Friedmann equation and the remaining mode gives a decelerating power law; at
$\alpha_{\rm v}^{(v)}$, the total gravitational density is constant and the
background is de Sitter. Likewise, $\alpha_{\rm s}^{(v)}$ yields an exactly
stiff-scaling background, while $\alpha_{\rm s}^{(c)}=-1/2$ cancels the stiff
pair and gives an accelerated power law. These are exact cancellation
identities within the declared closure, not generic predictions of EMSG. The
dust--vacuum example also makes the singular limit explicit:
$\rho_{\rm CDM}\propto a^{-9/2}$ for every $\alpha\ne0$. In the
radiation--vacuum example, $\rho_{\rm r}\propto a^{-(3+\sqrt3)}$, but the
positive-vacuum branch terminates at finite past redshift.
Reaching the benchmark $z_{\rm rec}=1100$ without crossing
$\rho_{\rm v}=0$ requires%

$\rho_{\rm v0}/\rho_{\rm r0}\gtrsim3.8\times10^{13}$.
For an order-unity present-day conventional vacuum density parameter and
order-unity gravitational weights, with the usual matter and photon normalization,
this bound makes the interacting radiation-like component negligible at
that epoch; it does not imply negligibility for arbitrary normalizations
or near-canceling gravitational weights.

For finite distinct $w_i\in(-1,1)$, both density eigenvectors have
opposite-sign components. Any positive initial pair therefore excites both
modes and has a maximal density-positive interval of the fixed logarithmic
width~\eqref{eq:positive-domain-width}. Its finite endpoints depend on the
initial ratio, but its width does not depend on that ratio or $\alpha$.
A zero of $H^2$ may shorten the cosmologically admissible interval. In
this domain single-mode positive density initial data are impossible, so
constant-$q$ histories require gravitational rank loss.%

For positive present-day conventional densities and $-1<w<1$, the vacuum
families reach $\rho_{\rm v}=0$ at a finite past scale factor and the stiff
families reach $\rho_{\rm s}=0$ at a finite future scale factor. Their
power-law, de Sitter, and stiff-like histories are therefore generally local
portions of the branches on which both conventional densities remain positive.
Continuation through either endpoint requires a new piecewise derivation
because $f\propto|\rho|$ is nondifferentiable at $\rho=0$.

The distinction between two-mode and one-mode expansion is also essential.
The general deceleration parameter~\eqref{eq:q-two-mode} evolves and can
change sign subject to the amplitude and positive-density conditions.
The dust--vacuum example realizes an admissible transition for
$-1/3<\alpha<1/2$, $\alpha\ne0$, but its branch endpoint and nonstandard
matter dilution preclude a complete cosmological history. The
modal-cancellation power laws have constant $q$ and cannot by themselves
provide the full decelerating-to-accelerating history.%
 The special solutions
are therefore local consistency diagnostics, not realistic cosmologies
established by this analysis. Neither the previously known linear two-fluid
modes nor the elementary power laws are claimed as new; the contribution is
their action-versus-closure classification, the qualified singular-limit
analysis, the rank-loss interpretation and global discriminant proof, and
the determination of physical positive-density domains.%

The principal outcome is consequently a consistency map for a singular
phenomenological closure: it identifies its exact backgrounds, its
rank-degenerate subfamilies, and the domains on which its fluid interpretation
is valid. It does not establish a microscopic interacting dark sector or an
observational solution to cosmological tensions. Existing scale-independent
EMSG background and perturbation analyses
\cite{Akarsu:2018aro,Fu:2024cjj,Dunsby:2025ahd,delaCruz-Dombriz:2026xsl}
adopt different effective-fluid and matter-conservation prescriptions; their
perturbation systems therefore cannot be transferred unchanged to the present
sector-separated closure. A microscopic EMSG completion would require an
explicit off-shell fluid action whose metric and matter variations reproduce
both the adopted contracted-Hessian prescription and the desired constituent
currents. A covariant phenomenological completion is logically possible
without such an action, but it would instead have to specify the transfer
four-vectors for all retained sectors, including their momentum components and
perturbations, together with pressure, entropy, and other closure relations.
A joint assessment of hyperbolicity, stability, and observational viability
using both background and perturbation observables requires such a
completion. Background consistency and kinematic limitations can already be
assessed at the level studied here.%

 Generalizations to unequal couplings or
other powers of $T_{\mu\nu}T^{\mu\nu}$ must likewise be formulated at the
chosen action or closure level. Until then, the exact solutions delimit the
consequences of the stated background closure; they are not evidence for a
microscopic interacting dark sector or a resolution of cosmological tensions.

\begin{acknowledgments}

\"{O}.A. acknowledges support from the Turkish Academy of
Sciences through the Outstanding Young Scientist Award (T\"{U}BA-GEB\.{I}P). \"{O}.A.
and N.K. are supported in part by T\"{U}B\.{I}TAK grant 122F124. N.K. acknowledges
support from Do\u{g}u\c{s} University Scientific Research Project 2021-22-D1-B01. This article is based upon work from COST Action CA21136,
``Addressing observational tensions in cosmology with systematics and
fundamental physics'' (CosmoVerse), supported by COST (European Cooperation in
Science and Technology). During the preparation and revision of this work, the authors used
the AI tools ChatGPT and Claude to assist with drafting and language
editing, cross-checking selected algebraic calculations, and developing
numerical verification code. All technical derivations, physical results,
and scientific conclusions were independently verified by the authors,
who reviewed and finalized the manuscript and take full responsibility
for its content.
\end{acknowledgments}

\appendix
\section{Coefficient-matrix degeneracy and density linear dependence}
\label{app:LD}

We first rewrite Eq.~\eqref{eq:wrons}, the total continuity equation, in matrix notation as
\begin{equation}
\dot{\Vec{\rho}}^{T}\Vec{x}_A+\Vec{\rho}^{T}\Vec{x}_B=0,
\label{eq:contmat}
\end{equation}
where
\begin{equation}
\begin{aligned}
\Vec{x}_A&=\begin{bmatrix}K_1\\K_2\end{bmatrix},&
\Vec{x}_B&=3H\begin{bmatrix}R_1\\R_2\end{bmatrix},\\
R_i&\equiv(1+w_i)(1+\alpha B_i),&
K_i&\equiv1+\alpha A_i.
\end{aligned}
\end{equation}
Using these vectors, we define
\begin{equation}
\mathbf{L}=\mathbf{W}\mathbf{X}=\begin{bmatrix}
\Vec{\rho}^{T}\Vec{x}_B & \Vec{\rho}^{T}\Vec{x}_A\\
\dot{\Vec{\rho}}^{T}\Vec{x}_B & \dot{\Vec{\rho}}^{T}\Vec{x}_A
\end{bmatrix},
\end{equation}
where $\mathbf{X}$ is the coefficient matrix
\begin{equation}
\label{eq:WXm}
\mathbf{X}=\begin{bmatrix}
\Vec{x}_B & \Vec{x}_A
\end{bmatrix}=\begin{bmatrix}
3HR_1 & K_1\\
3HR_2 & K_2
\end{bmatrix},
\end{equation}
with $H\neq0$, and $\mathbf{W}$ is the Wronskian matrix
\begin{equation}
\label{eq:Wm}
\mathbf{W}=\begin{bmatrix}
\Vec{\rho}^{T}\\
\dot{\Vec{\rho}}^{T}
\end{bmatrix}=\begin{bmatrix}
\rho_1 & \rho_2\\
\dot{\rho}_1 & \dot{\rho}_2
\end{bmatrix}.
\end{equation}
The condition $\operatorname{Tr}\mathbf{L}=0$ reproduces Eq.~\eqref{eq:contmat}. Moreover,
\begin{equation}
\det\mathbf{L}=\det\mathbf{W}\,\det\mathbf{X}.
\end{equation}
Hence, if $\det\mathbf{L}\neq0$, both $\mathbf{W}$ and $\mathbf{X}$ are nonsingular, and the nonvanishing Wronskian implies that the two energy densities are locally linearly independent. If $\det\mathbf{L}=0$, however, either $\det\mathbf{W}=0$ or $\det\mathbf{X}=0$, or both. Therefore, the singularity of $\mathbf{L}$ alone does not imply linear dependence of the energy densities.

The two determinants are
\begin{align}
\det\mathbf{W} &=\rho_1\dot{\rho}_2-\rho_2\dot{\rho}_1,
\label{eq:detw}\\
\det\mathbf{X}&=3H\Big[
(1+w_1)(1+\alpha B_1)(1+\alpha A_2)\nonumber\\
&\qquad\quad-(1+w_2)(1+\alpha B_2)(1+\alpha A_1)\Big]\nonumber\\
&=3H\Big[(w_1-w_2)+E(w_1,w_2)\Big].
\label{eq:detx}
\end{align}
The following Abel-identity argument is restricted to $w_1\neq w_2$, which is
the domain in which Eqs.~\eqref{eq:Q1}--\eqref{eq:Q2} and
\eqref{eq:dr1}--\eqref{eq:dr2} were obtained. The equal-EoS case must instead
be analyzed directly from Eqs.~\eqref{eq:cont11} and \eqref{eq:cont3}.
To make the linear-dependence statement precise, set $x=\ln a$ and define
\begin{equation}
\widehat W\equiv\rho_1\rho_2'-\rho_2\rho_1'.
\end{equation}
Because both densities satisfy Eqs.~\eqref{eq:dr1}--\eqref{eq:dr2}, Abel's
identity gives
\begin{equation}
\widehat W'=-C\widehat W.
\end{equation}
On a connected interval with $H\neq0$, one has
$\det\mathbf W=H\widehat W$. Hence a zero of $\det\mathbf W$ at one point
forces it to vanish throughout that interval. Wherever $\rho_2\neq0$, this is
equivalent to
\begin{equation}
\frac{\mathrm{d}}{\mathrm{d}t}\left(\frac{\rho_1}{\rho_2}\right)=0,
\end{equation}
and therefore $\rho_1=k\rho_2$ there, with constant $k$. By contrast,
$\det\mathbf{X}=0$ imposes the algebraic condition
\begin{equation}
(w_1-w_2)+E(w_1,w_2)=0,
\end{equation}
which is the rank-degeneracy condition introduced in
Sect.~\ref{sec:alphaparticular}. It is not an additional factor or dynamical
branch of the Li\'enard equation~\eqref{eq:mstro}, and it does not imply that
$\rho_1$ and $\rho_2$ are linearly dependent.
Without the EMSF correction, $E(w_1,w_2)=0$ and
$\det\mathbf X=0$ reduces to $w_1=w_2$. In the present model the EMSF terms
permit $\det\mathbf X=0$ also for $w_1\neq w_2$ at the two quadratic roots of
Eq.~\eqref{eq:aalp} when $s\neq0$, or at the unique linear root of
Eq.~\eqref{eq:a0} when $s=0$. For generic $\alpha$, the two gravitational equations
reconstruct the densities uniquely. At a special root that reconstruction is
rank deficient and must be performed in the original system. In both cases,
however, Eq.~\eqref{eq:mstro} follows independently from the density evolution.
Equations~\eqref{eq:edpt3}--\eqref{eq:edpt4} translate any special $\alpha$
value into EMSF-to-conventional density ratios. Equation~\eqref{eq:Oaalp}
gives those ratios only for the genuinely quadratic case; the linear case must
instead use Eq.~\eqref{eq:a0}. Within this unequal-EoS domain, linear dependence
of the two density solutions occurs precisely when $\det\mathbf W$ vanishes on
the connected interval with $H\neq0$ under consideration. 

The two conditions $\det\mathbf W=0$ and $\det\mathbf X=0$ each leave a
single mode in $H^2$, but for different reasons.
When $\det\mathbf W=0$, the two densities themselves lie on one mode,
$a^{-\gamma_+}$ or $a^{-\gamma_-}$; this is a restriction on the initial data
and is independent of $\alpha$. For distinct $w_i\in(-1,1)$ such
single-mode data necessarily have $\rho_1\rho_2<0$, by the proposition
in Sect.~\ref{sec:nvp}. When $\det\mathbf X=0$, both modes are generically
present in the densities, but one cancels in the gravitational%

combination $K_1\rho_1+K_2\rho_2$. This restricts $\alpha$ independently of
the initial density ratio; the resulting background must still satisfy
the physical-domain conditions. The constant-$q$ solutions of
Sect.~\ref{sec:alphaparticular} are of the second kind.%

\renewcommand{\bibfont}{\footnotesize}

\bibliography{bibliography}
\end{document}